\documentclass[aps,pra,reprint,amsmath,amssymb,
superscriptaddress,showpacs,showkeys,nofootinbib]{revtex4-2}
\usepackage{amsmath}    
\numberwithin{equation}{section}

\usepackage{graphicx}   
\usepackage{verbatim}   
\usepackage{color}      
\usepackage{xcolor}
\usepackage{subfigure}  
\usepackage{booktabs}
\usepackage{balance}
\usepackage[normalem]{ulem} 
\let\isout\sout \renewcommand{\sout}[1]{\ifmmode\text{\isout{\ensuremath{#1}}}\else\isout{#1}\fi}
\usepackage[utf8]{inputenc}
\usepackage[english]{babel}
\usepackage{dcolumn}    
\usepackage{bm}         
\usepackage{mathtools}
\usepackage{cancel}     
\newcommand{\ii}{\mathrm{i}}
\newcommand{\ee}{\mathrm{e}}

\newcommand{\eps}{\varepsilon}
\newcommand{\Hn}[1]{H_{#1}^{(1)}}
\newcommand{\Jn}[1]{J_{#1}}
\newcommand{\Ree}{\operatorname{Re}}
\newcommand{\Imm}{\operatorname{Im}}

\newcommand{\vect}[1]{\bm{#1}}

\newcommand{\QQI}{\mathbb{Q}(\ii)}
\newcommand{\diag}{\operatorname{diag}}

\DeclarePairedDelimiter{\norm}{\lVert}{\rVert}
\definecolor{vs}{rgb}{0.1,0.4,0.1}                  

\definecolor{remcyan}{rgb}{0.0,0.50,0.62}             
\usepackage{hyperref}   
\begin{document}

\title{High-Accuracy Semi-Analytical Method for Solving the Problem of Electromagnetic Wave Scattering by Arbitrary Ensembles of Parallel Circular Cylinders}
\author{V.V. Ternovski}
\email[]{\mbox{E-mail: vladimir.ternovskii\_at\_gmail.com}}
\affiliation{Lomonosov Moscow State University, Faculty of Computational Mathematics and Cybernetics, Department of Computational Methods,
Leninskie Gory 1, {bldg.~52,} 2nd Education Building, Moscow, 119991, Russia}
\author{M.I. Tribelsky}
\email[]{\mbox{E-mail: mitribel\_at\_gmail.com}}
\affiliation{Lomonosov Moscow State University, Faculty of Physics,
Department of Polymer and Crystal Physics,
Leninskie Gory 1, bldg.~2, Moscow, 119991, Russia}
\affiliation{Center for Photonics and 2D Materials, Moscow Institute of Physics and Technology (MIPT), 141700 Dolgoprudny, Russia}
\date{\today}

\begin{abstract}
A method is proposed for solving the two-dimensional problem of electromagnetic wave scattering by a cluster of an arbitrary number of parallel, infinitely long, homogeneous, non-overlapping right circular cylinders. The cylinders may have arbitrary radii and complex permittivities, and their axes, while remaining parallel, may occupy arbitrary positions in the transverse plane. The solution is constructed using an analytical expansion of the electromagnetic field in cylindrical harmonics. Multiple scattering is taken into account by Graf's addition theorem, which leads to a system of linear equations for the expansion coefficients. This system is solved numerically with condition number monitoring and, when necessary, extended-precision arithmetic, followed by a multistage verification of convergence. The method provides numerically verified solutions with controlled accuracy over a wide range of parameters, including densely packed subwavelength configurations. As an example, scattering of a normally incident, linearly polarized monochromatic plane wave by a subwavelength cluster of three identical aluminum nanocylinders (nanowires) is studied. The scattering, absorption, and extinction cross sections, as well as the scattering indicatrix, are computed and analyzed. Streamlines of the Poynting vector field are constructed, demonstrating redistribution of the energy flux between the cylinders of the cluster and the formation of localized regions of field enhancement near their surfaces.
\end{abstract}
\maketitle

\noindent
{\it Keywords:} multiple scattering, Mie theory, cylindrical
harmonics, addition theorem, trimer, field enhancement, Poynting vector, optical theorem,
plasmonics, Gaussian rational numbers, exact arithmetic.

\section{Introduction\label{sec:intro}}

Scattering of electromagnetic waves by ensembles of parallel cylinders
is a classical problem in electrodynamics. In addition to the purely academic interest it is important to design photonic crystals, plasmonic nanostructures, optical antennas, and 
metamaterials~\cite{bohren1983,tsang2000,rahmani2004}. For a single
cylinder, an exact solution is available in the framework of the Mie multipole expansion.

For clusters of two or more elements, simple analytical solutions are
generally unavailable, so numerical methods become essential. Numerous
freely available and commercial packages exist for this class of problems.
Without discussing their features in detail, we note that any finite-difference
scheme inevitably approximates both the cylinder surfaces and the structure
of the electromagnetic field. Therefore, when computing near fields in the most
application-relevant cases, where the field may have sharp spatial variations,
such schemes either cannot provide the required accuracy or imply extremely
small-step discretization, which makes the computations very costly. On top of that, in these cases, the question about convergence of the numerical scheme arises. 

In the present work, we propose an alternative semi-analytical approach
that addresses this difficulty. The method is based on expanding the field in
cylindrical harmonics about the axis of each cylinder and applying Graf's addition
theorem for Bessel and Hankel functions. The resulting self-consistent system
of linear equations for the modal coefficients is solved numerically by an original
technique designed to ensure high accuracy.

Similar approaches have already been applied to the class of problems
considered here~\cite{twersky1952,linton2005,martin2006,schaefer2012calculation,beutel2024treams,2D_Loulas2025,tanaka2026calculating}. A major drawback of these methods in modeling resonant and densely packed structures is the ill-conditioning of the resulting linear systems, which can severely degrade the numerical accuracy and slow iterative convergence. The multistage adaptive solver developed in the present work is designed specifically to overcome this difficulty.

We emphasize, however, that despite its high accuracy and the other advantages discussed below, the proposed method is applicable only to a relatively limited class of problems similar to those considered here. Its generalization to other situations, with the exception of scattering by ensembles of spheres~\cite{mackowski1996}, involves considerable difficulties. Therefore, it cannot replace existing methods with a broader domain of applicability; rather, it can play an important role in testing and validating them.

As an example, we consider the scattering of a normally incident, linearly polarized monochromatic plane wave by a symmetric trimer: a cluster of three identical aluminum nanocylinders (nanowires) of circular cross section whose axes pass through the vertices of an equilateral triangle lying in the plane perpendicular to these axes. This choice is motivated by the fact that aluminum is a promising plasmonic material owing to its low dissipative losses in the UV spectral range~\cite{knight2014,gerard2015,thogersen2023plasmonic}. The symmetric trimer, in turn, has a special place among nano-oligomers: despite its relatively simple geometry, it has a rich structure of collective modes that are sensitive to the distance between the cylinders and to their radii~\cite{nordlander2004,prodan2003,brandl2006,alegret2008,Zhao2024}.

The paper is organized as follows. Section~\ref{sec:formulation} presents the derivation of the vector Helmholtz equation and its reduction to a two-dimensional scalar problem. The analytical treatment of the problem for an arbitrary ensemble of cylinders, taking into account multiple scattering, is given in Sec.~\ref{sec:expansions}. Section~\ref{sec:summary} briefly summarizes the main analytical results. The numerical algorithm, including the four-level adaptive solver for the truncated system of linear equations for the modal coefficients, is described in Sec.~\ref{sec:numerics}. The formulas for computing the Poynting vector field and its streamlines are given in Sec.~\ref{sec:poynting}. Numerical results for the aluminum trimer are presented in Sec.~\ref{sec:results}, and the conclusions are given in Sec.~\ref{sec:conclusion}.

\section{Formulation of the problem
\label{sec:formulation}}

\subsection{Preliminary remarks}

We consider the scattering of a linearly polarized monochromatic plane wave propagating in vacuum at normal incidence on an ensemble of infinitely long parallel cylinders of circular cross section. Under these assumptions, the problem is two-dimensional. Since plane waves form a complete set of functions, an arbitrary laser pulse can be expanded in them. Owing to the linearity of the problem with respect to the fields $\mathbf{E}$ and $\mathbf{H}$, each monochromatic component can be considered independently. The Poynting vector field $\mathbf{S}$ is then computed from the known total fields $\mathbf{E}$ and $\mathbf{H}$.

Thus, the analysis of scattering of a monochromatic plane wave is fundamental: solutions for more complex incident fields can be obtained as superpositions of such waves, provided material dispersion is taken into account. Such generalizations are not considered in the present work.

\vspace*{4pt}
\subsection{Vector Helmholtz equations
\label{sec:governing}}
\vspace*{4pt}

We use the Gaussian system of units, in which the fields $\mathbf{E}$ and $\mathbf{H}$ have the same dimension, and the permittivity and permeability of vacuum are equal to unity. The cylinders are assumed to be isotropic and spatially homogeneous. In general, they may have different complex permittivities $\eps_p = \eps_p'+ \ii\eps_p''$, where $p$ is the cylinder index. Owing to this spatial homogeneity, $\eps_p$ is independent of the coordinates inside each cylinder. At optical frequencies, the magnetic permeability is equal to unity~\cite{LL_Electrodyn}, which will be assumed in what follows. For time-harmonic fields with the time dependence
$\ee^{-\ii\omega t}$, the source-free Maxwell equations read
\begin{equation}
\nabla\times\vect{E}= \ii k_0\,\vect{H},\qquad
\nabla\times\vect{H}= -\ii k_0\,\eps_{(j)p}\,\vect{E}.
\label{eq:maxwell}
\end{equation}
Here $\eps_{(j)p}$ is equal to $\eps_p$ inside the cylinders and to unity outside them; $k_0=\omega/c$, where $c$ is the speed of light in a vacuum.

In the absence of free charges and currents, the conditions
$\nabla\cdot(\eps_{(j)p}\vect{E})=0$ and $\nabla\cdot\vect{H}=0$ hold. Within each homogeneous region, where $\eps_{(j)p}=\eps_{p}=\mathrm{const}$, they reduce to \mbox{$\nabla\cdot\vect{E}=0$} and $\nabla\cdot\vect{H}=0$.

Applying the curl operation to Eq.~\eqref{eq:maxwell} and using the identity
\mbox{$\nabla\times(\nabla\times\vect{F})=\nabla(\nabla\cdot\vect{F})-\nabla^2\vect{F}$,} we obtain in the standard way the vector Helmholtz equations~\cite{LL_Electrodyn}
\begin{equation*}
\nabla^2\vect{E}+k_{(j)p}^2\vect{E}=\vect{0},\qquad
\nabla^2\vect{H}+k_{(j)p}^2\vect{H}=\vect{0},
\end{equation*}
\vspace*{4pt}
where $k_{(j)p}=k_0\sqrt{\eps_{(j)p}}$ is the local wavenumber. The branch is chosen so that, for passive media, $\Imm(k_{(j)p})\ge 0$.

\subsection{Dimensionless variables
\label{sec:dimless}}

To make all subsequent relations independent of the system of units (SI, Gaussian, etc.), it is convenient to introduce dimensionless field variables by normalizing all fields to the amplitudes of the corresponding fields in the incident wave. Since only such dimensionless variables will be used below, we keep the same notation as for the dimensional quantities; this should not cause confusion. The time-averaged Poynting vector $\vect{S}$ is normalized to its magnitude in the incident wave. This normalization fixes the overall scale and makes it possible, when needed, to compare the fields $\vect{E}$, $\vect{H}$, and the energy flux $\vect{S}$ in the same plot.

\subsection{Geometry of the problem. Scalar wave equation\label{sec:geometry}}

The cylinders are assumed to be non-overlapping and non-touching. The case of touching cylinders, often romantically called ``kissing'' cylinders~\cite{lei2010broadband}, requires separate treatment. We do not consider it here because, for the present problem, this question is mainly of academic interest. Although the approach developed below is formally applicable only for finite gaps between the cylinders, the threshold gap values are so small that they impose no significant restrictions on its practical application. For example, in nanoclusters the approach remains applicable down to gaps of atomic size, which, within the macroscopic description used here, is effectively equivalent to touching cylinders.

Let us choose a coordinate system with the $z$ axis parallel to the cylinder axes. As noted above, we consider only normal incidence with respect to the cylinder axes, so that the $z$ component of the incident wave vector $\mathbf{k}_0$ is zero. As for the orientation of the polarization plane, owing to the linearity of Maxwell's equations, the fields $\mathbf{E}$ and $\mathbf{H}$ for an arbitrary orientation of this plane can be represented as a linear combination of waves with the vector $\mathbf{E}$ parallel and perpendicular to the $z$ axis, with the corresponding orientation of the vector $\mathbf{H}$.

Thus, without loss of generality, it is sufficient to consider only these two independent polarizations~\cite{Tribelsky2023}. They are usually called TM and TE. Here, however, we use the terms $E_z$~polarization and $H_z$~polarization to emphasize which field is oriented along the $z$ axis.

Under the assumptions made above, the electromagnetic fields are independent of the coordinate $z$. Moreover, by symmetry, for the $E_z$~polarization, not only the incident wave but also the scattered wave and the field inside the cylinders have only the $z$ component of the electric field different from zero. This single component satisfies the wave equation
\begin{equation}
\bigl(\nabla_\perp^2 + k_{(j)p}^2\bigr)\,E_z = 0,
\label{eq:helmholtz_psi_TM}
\end{equation}
where \(\nabla_\perp^2 = \partial^2/\partial x^2 + \partial^2/\partial y^2\) is the two-dimensional Laplace operator.

The transverse components of the magnetic field are expressed in terms of \(E_z\) according to Maxwell's equations, see Eq.~\eqref{eq:maxwell}, as
\begin{equation*}
H_x = \frac{c}{\mathrm{i}\omega}\frac{\partial E_z}{\partial y}, \qquad
H_y = -\frac{c}{\mathrm{i}\omega}\frac{\partial E_z}{\partial x}.
\end{equation*}

Similarly, for the $H_z$~polarization we have
\begin{equation}
\bigl(\nabla_\perp^2 + k_{(j)p}^2\bigr)\,H_z = 0,
\label{eq:helmholtz_psi_TE}
\end{equation}
\begin{equation}
E_x = -\frac{c}{\mathrm{i}\omega\,\varepsilon_{(j)p}}\frac{\partial H_z}{\partial y}, \qquad
E_y = \frac{c}{\mathrm{i}\omega\,\varepsilon_{(j)p}}\frac{\partial H_z}{\partial x}.
\label{eq:TE_transverse}
\end{equation}

It is convenient to combine Eqs.~\eqref{eq:helmholtz_psi_TM} and \eqref{eq:helmholtz_psi_TE} into a single scalar wave equation:
\begin{equation}
\bigl(\nabla_\perp^2 + k_{(j)p}^2\bigr)\,\psi = 0.
\label{eq:helmholtz_psi}
\end{equation}
Here
\[
\psi =
\begin{cases}
E_z, & \text{for the }E_z\text{~polarization},\\[4pt]
H_z, & \text{for the }H_z\text{~polarization}.
\end{cases}
\]

\subsection{Normalized Poynting vector\label{sec:poynting_formulas}}

In the dimensionless variables introduced above, the time-averaged Poynting vector is given by the following formulas:

\noindent
for the $E_z$~polarization,
\begin{equation}
\vect{S} = \frac{1}{k_0}\,\Imm\!\left(\psi^*\,\nabla\psi\right),
\label{eq:S_unified_Ez}
\end{equation}
for the $H_z$~polarization,
\begin{equation}
\vect{S} = \frac{1}{k_0}\,\Imm\!\left[\frac{1}{\eps_{(j)p}}\,\psi^*\,\nabla\psi\right].
\label{eq:S_unified_Hz}
\end{equation}

Formula~\eqref{eq:S_unified_Hz} contains the piecewise-constant function $\eps_{(j)p}$. In computations near the interfaces, one should use the local value of the permittivity in the region, outside or inside the cylinder, where the observation point is located. This ensures that the energy flux is computed correctly in each medium.

In expanded form, these expressions are written as
\begin{equation*}
S_x = {\frac{1}{k_0}}\,\Imm\!\left(E_z^{*}\,
      \frac{\partial E_z}{\partial x}\right),
\quad
S_y = {\frac{1}{k_0}}\,\Imm\!\left(E_z^{*}\,
      \frac{\partial E_z}{\partial y}\right),
\end{equation*}
for the $E_z$~polarization, and
\begin{equation*}
S_x = {\frac{1}{k_0}}\,\Imm\!\left(\frac{H_z^*}{\eps_{(j)p}}\,
      \frac{\partial H_z}{\partial x}\right),
\quad
S_y = {\frac{1}{k_0}}\,\Imm\!\left(\frac{H_z^*}{\eps_{(j)p}}\,
      \frac{\partial H_z}{\partial y}\right),
\end{equation*}
for the $H_z$~polarization.

\section{Analytical solution for an arbitrary ensemble of parallel non-overlapping circular cylinders
\label{sec:expansions}}
\vspace*{6pt}

In this section, following the approach developed in Ref.~\cite{schaefer2012calculation}, we present the exact solution of the problem formulated above for an arbitrary number of arbitrarily placed cylinders. We refer to the original coordinate system as the \emph{global} coordinate system. To apply the method, we need expansions of the function $\psi$ in cylindrical harmonics in local coordinate systems whose $z$ axes coincide with the axes of the individual cylinders.

Let $\vect{r}$ be the two-dimensional position vector of a point in the global coordinate system in the plane perpendicular to the $z$ axis. Let $\vect{r}_p$ be the position vector of {\it the same point\/} in the local coordinate system associated with the $p$-th cylinder, with polar coordinates $(r_p, \varphi_p)$ in that local system. Then,
\begin{equation*}
  \vect{r} = \vect{c}_p+\vect{r}_p,
\end{equation*}

\noindent
where $\vect{c}_p$ is the position vector of the center of the local system in the global one.

The main idea of the method is as follows. For each cylinder, in its local coordinate system, one can write a formally exact general solution of the wave equation, separately outside and inside the cylinder, as an infinite series in appropriate cylindrical functions with unknown coefficients. Outside the cylinders, the field is represented as the superposition of such a solution and the incident field.

The problem then reduces to finding these unknown coefficients. This is done by imposing the boundary conditions on the surface of each cylinder, which leads to a chain of linear equations for the coefficients. Solving this chain completes the analytical treatment of the problem.

Thus, the task is to construct the corresponding formal solution, determine the boundary conditions, and then implement the procedure in practice. We now proceed with this construction.

As noted above, the problem splits into two coupled boundary-value problems: the exterior problem, which determines the field outside the cylinders, and the interior problem, which describes the field inside them. These two problems are coupled through the boundary conditions on the cylinder surfaces. We discuss them separately.

\subsection{Exterior boundary-value problem}
\vspace*{-2pt}

The general solution of the wave equation associated with each cylinder and satisfying the outgoing-wave condition at infinity is written in terms of Hankel functions of the first kind $H_n^{(1)}$~\cite{bohren1983}. Thus, the total field scattered by all cylinders has the form
\begin{eqnarray}
   & & \!\!\!\!\!\!\!\!\!\!\! \psi^{(\rm sca)}(\vect{r}) =
\sum_{p=1}^P\psi^{(\rm sca)}_p(\vect{r})\label{eq:ext_field} \\
   & & \!\!\!\!\!\!\!\!\!\!\! \psi^{(\rm sca)}_p(\vect{r}) \equiv \sum_{n=-\infty}^{\infty} A_{np}\,\Hn{n}(k_0r_p)\,
\ee^{\ii n\varphi_p};\; r_p \geq a_p, \label{eq:psi_sca_p}
\end{eqnarray}
where $P$ is the number of cylinders in the cluster, $a_p$ is the radius of the $p$-th cylinder, and $A_{np}$ are yet unknown coefficients. We emphasize that $\vect{r}$ in Eq.~\eqref{eq:psi_sca_p} is the position vector of the observation point in the global coordinate system, whereas $(r_p,\varphi_p)$ are the coordinates of \emph{the same} point in \emph{each} of the local coordinate systems.
The total field in the exterior boundary-value problem, $\psi^{\rm (out)}$, is the sum of the incident field $\psi^{(0)}$ and the scattered field $\psi^{(\rm sca)}$:
\vspace*{4pt}
\begin{equation}\label{eq:psi_1}
  \psi^{\rm (out)} = \psi^{(0)} + \psi^{(\rm sca)}.
\end{equation}

Let us now pass to a fixed $p$-th local coordinate system and separate from $\psi^{\rm (out)}$ the field scattered by the cylinder associated with this coordinate system. In other words, in the exterior neighborhood of this cylinder, we represent $\psi^{\rm (out)}$ as

\begin{equation}\label{eq:psi_ext}
  \psi^{\rm (out)} = \psi^{(\rm ext)} + \psi_p^{(\rm sca)},
\end{equation}\\
where $\psi_p^{(\rm sca)}$ is described by Eq.~\eqref{eq:psi_sca_p}, and $\psi^{\rm (ext)}(\vect{r})$ is the field acting on the $p$-th cylinder from outside, i.e., the superposition of the incident field and the waves scattered by the remaining $P-1$ cylinders. We expand the fields $\psi^{(\rm ext)}$ and $\psi_p^{(\rm sca)}$ in cylindrical harmonics in the chosen local coordinate system. For $\psi^{(\rm ext)}$, this expansion is
\begin{equation}\label{eq:general_cyl_expantion_out}
  {\psi_p^{(\rm ext)}(\vect{r}) = \sum_{n=-\infty}^{\infty}  B_{np} \Jn{n}(k_0r_p)\,
\ee^{\ii n\varphi_p}},
\end{equation}
where the coefficients $B_{np}$ are determined during the solution of the problem.

For the incident plane wave propagating in the direction
$\vect{\kappa} = (\cos\varphi_{0},\,\sin\varphi_{0})$,
\begin{equation*}
\psi^{(0)}(\vect{r}) = \ee^{\ii k_0\,\vect{\kappa}\cdot\vect{r}}.
\end{equation*}
Its expansion in the form~\eqref{eq:general_cyl_expantion_out} follows from the Jacobi--Anger identity
\vspace*{-4pt}
\begin{equation*}
  \ee^{\ii\zeta\cos\gamma} \equiv \sum_{n=-\infty}^{\infty}\ii^nJ_n(\zeta)\ee^{\ii n\gamma}.
\end{equation*}
This gives
\begin{equation}
B_{np}^{(0)} = \ii^n\,\ee^{-\ii n\varphi_0}\,
\ee^{\ii k_0\,\vect{\kappa}\cdot\vect{c}_p}.
\label{eq:Binc}
\end{equation}
\vspace*{-24pt}
\subsection{Interior boundary-value problem}
\vspace*{-6pt}

The general solution of the wave equation inside each cylinder, finite on its axis, has the form~\cite{bohren1983}
\begin{equation}
\psi^{\rm (in)}_p(\vect{r}_p) = \sum_{n=-\infty}^{\infty} D_{np}\,\Jn{n}(k_pr_p)\,
\ee^{\ii n\varphi_p}; \quad r_p \leq a_p,
\label{eq:int_field}
\end{equation}
where $D_{np}$ are the unknown interior coefficients. Here $k_p \equiv k_0\sqrt{\eps_p}$.

Equations~\eqref{eq:psi_sca_p}, \eqref{eq:general_cyl_expantion_out}, and \eqref{eq:int_field} are the standard multipole expansions widely used in exact solutions of the diffraction problem for a monochromatic plane wave incident on a cylinder. Here $n=\pm 1$ corresponds to dipole radiation, $n=\pm 2$ to quadrupole radiation, and so on~\cite{bohren1983}.

\subsection{Boundary conditions on the cylinder surfaces}

The boundary conditions require continuity of the tangential components of the fields $\vect{E}$ and $\vect{H}$ on the surfaces of the cylinders~\cite{bohren1983}. It is convenient to write these conditions in the local cylindrical coordinate systems associated with the individual cylinders. We consider the two independent polarizations separately.

\emph{$E_z$ polarization}. The nonzero components are $E_z$, $H_{\varphi_p}$, and~$H_{r_p}$, and the scalar variable in Eq.~\eqref{eq:helmholtz_psi} is $E_z$. Since $E_z$ is tangential to the cylinder surface, one boundary condition reduces to its continuity: \mbox{$E_z\big|_{-}=E_z\big|_{+}$.}

The field $\vect{H}$ is computed from the known field $\vect{E}$ using the Maxwell equation \mbox{$\vect{H}=-(\ii/k_0)\nabla\times\vect{E}$.} Continuity of the component $H_\varphi$ then gives $\partial_{r_p}E_z\big|_{-}=\partial_{r_p}E_z\big|_{+}$.

\emph{$H_z$ polarization}. The nonzero components are $H_z$, $E_{\varphi_p}$, and~$E_{r_p}$, and the scalar variable is $H_z$. Continuity of $H_z$ gives $H_z\big|_{-}=H_z\big|_{+}$. The field $\vect{E}$ is computed from $\vect{H}$ according to {
\mbox{$\vect{E}=\displaystyle{\frac{\ii c}{\omega\eps_{(j)p}}}\nabla\times\vect{H}$}, see Eq.~\eqref{eq:TE_transverse}. Outside the cylinders $\eps_{(j)p}=1$, whereas inside them \mbox{$\eps_{(j)p}=\eps_p \neq 1$.}} Therefore, continuity of $E_\varphi$ gives \mbox{$(1/\eps_p)\,\partial_{r_p} H_z\big|_{-} = \partial_{r_p} H_z\big|_{+}$.}

Passing to the scalar variable $\psi$ and combining the two cases, the boundary conditions on the surface of each cylinder can be written as

\begin{equation}\label{eq:bc_psi}
  \psi\big|_{-}=\psi\big|_{+}, \qquad \alpha_p\left.\frac{\partial \psi}{\partial r_p}\right|_{-} = \left.\frac{\partial \psi}{\partial r_p}\right|_{+}.
\end{equation}

\mbox{}\\
\noindent Here, for the $E_z$~polarization, $\psi = E_z$ and $\alpha_p=1$, whereas, for the $H_z$~polarization, $\psi = H_z$ and $\alpha_p=1/\eps_p$. For clarity, we emphasize that while $\eps_{(j)p}$ introduced above is a piecewise-constant function of the coordinates, $\alpha_p$ is a coordinate-independent material constant.

\subsection{Scattering of an arbitrary linearly polarized monochromatic wave by a single cylinder
\label{sec:mie}}

In the problem considered here, each cylinder is placed in the incident field and in the field of waves multiply rescattered by all cylinders of the cluster. The latter field is completely determined by the coefficients $A_{np}$. Our task is to obtain the equations for these coefficients. To this end, we first solve the scattering problem for a single cylinder excited by an {\it arbitrary\/} monochromatic field of the form~\eqref{eq:general_cyl_expantion_out}, and then close the problem by relating this solution to the scattering problem for each cylinder of the cluster.

Thus, we first solve the scattering problem for the external field~\eqref{eq:general_cyl_expantion_out} without specifying the coefficients $B_{np}$. This problem is straightforward.

Let us choose and fix a value of $p$. The general solutions for the field scattered by this cylinder and for the field inside it are given by Eqs.~\eqref{eq:psi_sca_p} and \eqref{eq:int_field}, respectively. Applying the boundary conditions~\eqref{eq:bc_psi} and using the orthogonality of $\ee^{\ii n \varphi_p}$ for different $n$, we obtain
\begin{eqnarray}
  & &  \mbox{\hspace*{-7mm}}
 \!\!\! \Jn{n}\left(x_p^{(0)}\right)B_{np} + \Hn{n}\left(x_p^{(0)}\right)A_{np} =\Jn{n}\left(x_p^{(c)}\right)D_{np},\label{eq:bc1} \\
   & & \mbox{\hspace*{-7mm}}
   \!\!\!  \left[\Jn{n}'\left(x_p^{(0)}\right)B_{np} + {\Hn{n}}'\left(x_p^{(0)}\right)A_{np}\right]k_0 = \label{eq:bc2}\\
   & &  \mbox{\hspace*{38mm}}=\Jn{n}'\left(x_p^{(c)}\right)D_{np}\alpha_p k_p, \nonumber
\end{eqnarray}
where the notation $x_p^{(0)} \equiv k_0 a_p$ has been introduced (this quantity is often called the \emph{size parameter}, a name we also use), $x_p^{(c)} \equiv k_p a_p\equiv k_0a_p\sqrt{\varepsilon_p}$, and the prime denotes differentiation with respect to the full argument of the function.

Equations~\eqref{eq:bc1} and \eqref{eq:bc2} may be rewritten as

\begin{eqnarray}
   & & A_{np} = s_{np}\,B_{np}, \label{eq:Mie_diagonal_A} \\
   & & D_{np} = t_{np}\,B_{np}. \label{eq:Mie_diagonal_D}
\end{eqnarray}

\noindent
Here the \emph{scattering coefficients}\footnote{The scattering coefficients used here differ in sign from the coefficients usually employed in such problems~\cite{bohren1983}.} ($s_{np}$) and the \emph{transmission coefficients} ($t_{np}$), which depend on the polarization through the parameter $\alpha_p$, are defined by the expressions

\begin{eqnarray}
  & &\hspace*{-4mm} s_{np} \!=\! \label{eq:s}\\
  & &\hspace*{-4mm} \!=\! \frac{\alpha_p k_p\Jn{n}'\left(\!x_p^{(c)}\!\right)\!\Jn{n}\left(\!x_p^{(0)}\!\right)\! - \!
  k_0\Jn{n}'\left(\!x_p^{(0)}\!\right)\!\Jn{n}\left(\!x_p^{(c)}\!\right)}{k_0\Jn{n}\left(\!x_p^{(c)}\!\right)\!{\Hn{n}}'\left(\!x_p^{(0)}\!\right) - \! \alpha_p k_p\Jn{n}'\left(\!x_p^{(c)}\!\right)\!\Hn{n}\left(\!x_p^{(0)}\!\right) },\nonumber
\\
   & &\hspace*{-4mm} t_{np} \!=\! \label{eq:t}\\
   & &\hspace*{-4mm} \!=\! \frac{2\ii k_0}{\pi x_p^{(0)}\!\!\left[k_0\Jn{n}\left(\!x_p^{(c)}\!\right)\!{\Hn{n}}'\left(\!x_p^{(0)}\!\right)\! - \! \alpha_p k_p\Jn{n}'\left(\!x_p^{(c)}\!\right)\!\Hn{n}\left(\!x_p^{(0)}\!\right)\right]},\nonumber
\end{eqnarray}
and, in deriving Eq.~\eqref{eq:t}, we used the identity
\[
\Jn{n}(z) {{\Hn{n}}'(z)}-{\Jn{n}}'(z)\Hn{n}(z) \equiv \frac{2\ii}{\pi z}.
\]

\subsection{Graf's addition theorem. Translation matrix
\label{sec:multiple}}
To complete the problem formulation, it remains to express the coefficients $B_{np}$ in Eqs.~\eqref{eq:Mie_diagonal_A} and \eqref{eq:Mie_diagonal_D} in terms of the known coefficients $B_{np}^{(0)}$, see Eq.~\eqref{eq:Binc}, and the set of coefficients $A_{nq}$. This is done using Graf's addition theorem~\cite{abramowitz1964,watson1944}, which represents the field of an outgoing wave centered at point $q$ as an expansion in Bessel functions centered at point $p$.

To apply this theorem, choose two arbitrary cylinders $p$ and $q$ from the cluster. Let the positions of their axes in the global coordinate system be given by the two-dimensional vectors $\vect{c}_p$ and $\vect{c}_q$. Define the vector $\vect{R}_{pq} = \vect{c}_p - \vect{c}_q$, which in the local coordinate system associated with cylinder $q$ has polar coordinates\footnote{For clarity, we denote the polar angles of the coordinates of the cylinder axes by $\theta$, whereas all other polar angles are \mbox{denoted by $\varphi$.}} $(R_{pq}, \theta_{pq})$.

Consider the $m$-th partial scattered wave associated with cylinder $q$. In the chosen local coordinate system, this field is described by $\Hn{m}(k_0r_q)\,\ee^{\ii m\varphi_q}$, see Eq.~\eqref{eq:psi_sca_p}. According to Graf's theorem,

\begin{eqnarray}
   & & \Hn{m}(k_0r_q)\,\ee^{\ii m\varphi_q}
= \label{eq:graf} \\
   & & =\sum_{n=-\infty}^{\infty}\underbrace{
\Hn{m-n}(k_0 R_{pq})\,\ee^{\ii (m-n)\theta_{pq}}
}_{(T_{pq})_{nm}} \;\Jn{n}(k_0 r_p)\,\ee^{\ii n\varphi_p}.
\nonumber
\end{eqnarray}
Formula~\eqref{eq:graf} is valid under the condition $r_p < R_{pq}$, which is always satisfied in the vicinity of the surface of cylinder $p$ if the cylinders in the cluster do not intersect.

Equation~\eqref{eq:graf} defines the elements of the \emph{translation matrix}:
\begin{equation}
(T_{pq})_{nm} = \Hn{m-n}(k_0 R_{pq})\,\ee^{\ii(m-n)\theta_{pq}}.
\label{eq:T_pq}
\end{equation}
It is essential that the right-hand side of Eq.~\eqref{eq:graf} contains
\emph{regular} Bessel functions $\Jn{n}(k_0r_p)$ rather than singular Hankel
functions. The latter enter this expansion only as coefficients at fixed finite arguments. Therefore, the field scattered by cylinder~$q$, re-expanded about the center of cylinder~$p$, has the same functional form as the expansion~\eqref{eq:general_cyl_expantion_out}. This makes it possible to combine algebraically the contributions of the incident plane wave and of the cylindrical waves scattered by the cylinders of the cluster. Replacing the dummy index $n$ by $m$ in Eq.~\eqref{eq:psi_sca_p}, substituting Eq.~\eqref{eq:graf}, summing over $q$, and adding the incident wave in the representation~\eqref{eq:general_cyl_expantion_out}, \eqref{eq:Binc}, we obtain the following expression for the external field near the $p$-th cylinder:
\begin{eqnarray*}
   & &\!\!\!\!\! \psi_p^{(\mathrm{ext})}(\mathbf r_p)
= \\
   & & \!\!\!\!\!=\sum_{n=-\infty}^{\infty}
\left[
B_{np}^{(0)}
+
\mathop{{\sum}'}_{q=1}\limits^{\raisebox{-2pt}{\scriptsize{$P$}}}
\sum_{m=-\infty}^{\infty}
(T_{pq})_{nm}\,A_{mq}
\right]
J_n(k_0r_p)\,e^{\mathrm{i} n\varphi_p}.\nonumber
\end{eqnarray*}
Here the prime at the summation sign means that $q\neq p$. Comparing this formula with the expansion~\eqref{eq:general_cyl_expantion_out} gives an explicit expression for the coefficients $B_{np}$:
\begin{equation}\label{eq:Bpn}
  B_{np} =  \mathop{{\sum}'}_{q=1}\limits^{\raisebox{-2pt}{\scriptsize{$P$}}}
  \sum_{m=-\infty}^{\infty} A_{mq}\,(T_{pq})_{nm} + B_{np}^{(0)}.
\end{equation}

Substituting these values into Eq.~\eqref{eq:Mie_diagonal_A}, we obtain a closed infinite system of linear equations, valid for each value of $p$---that is, $P$ such systems in total:

\begin{equation}
A_{np} - s_{np}\mathop{{\sum}'}_{q=1}^{P}\;\sum_{m=-\infty}^{\infty}
(T_{pq})_{nm}\;A_{mq}
= s_{np}\,B_{np}^{(0)}.
\label{eq:self_consistent}
\end{equation}
\noindent
The solutions of Eqs.~\eqref{eq:self_consistent}, together with Eqs.~\eqref{eq:general_cyl_expantion_out}, \eqref{eq:int_field}, \eqref{eq:Mie_diagonal_A}, \eqref{eq:Mie_diagonal_D}, and \eqref{eq:Bpn}, constitute the exact analytical solution of the problem in the form of infinite series.

\subsection{Far field. Scattering amplitude, cross-sections, and the optical theorem\label{sec:opt_theorem}}

In the far field, the wave scattered by any object whose properties are independent of the coordinate $z$ and whose transverse cross-section is finite is an outgoing cylindrical wave. In the global coordinate system, such a wave is described by the well-known asymptotic form of $H^{(1)}_n(z)$ as $z \to \infty$:
\begin{equation}
\Hn{n}(z)\simeq
\sqrt{\frac{2}{\pi z}}\;
\ee^{\ii(z - n\pi/2 - \pi/4)},
\label{eq:hankel_asymptotic}
\end{equation}
and can be represented as~\cite{Hulst2018}
\begin{equation}
\psi_{\rm sca}(r,\varphi)\;\simeq\;
f(\varphi)\,\sqrt{\frac{2}{\pi k_0 r}}\;
\ee^{\ii(k_0 r - \pi/4)},
\quad r\to\infty.
\label{eq:far_field_amplitude}
\end{equation}
The function $f(\varphi)$ is the \emph{scattering amplitude}. It follows from Eq.~\eqref{eq:far_field_amplitude} that $f(\varphi)$ is dimensionless.

The scattering ($C_{\rm sca}$), absorption ($C_{\rm abs}$), and extinction ($C_{\rm ext}\equiv C_{\rm sca}+C_{\rm abs}$) cross-sections are expressed in the standard way through the scattering amplitude~\cite{Hulst2018}. Before computing them, however, we recall that, in the presence of continuous translational symmetry along the $z$ axis, all cross-sections are computed per unit length along this axis. Therefore, in the present two-dimensional problem, $C_{\rm ext,\,sca,\,abs}$ have the dimension of length rather than area.

To compute $C_{\rm sca}$, it is sufficient to calculate the flux of the Poynting vector through a circle of large radius $r$ centered at the origin of the global coordinate system. Substituting Eq.~\eqref{eq:far_field_amplitude} into Eqs.~\eqref{eq:S_unified_Ez} and \eqref{eq:S_unified_Hz}, which are identical in the far field because $\eps_{(j)p} = 1$ there, gives the radial component of the Poynting vector of the scattered wave:
\begin{equation*}
  S^{({\rm sca})}_r \simeq \frac{2|f(\varphi)|^2}{\pi r{ k_0}}, \quad r \to \infty.
\end{equation*}
Integration over the circle gives the scattering cross-section~\cite{Hulst2018}:

\begin{equation}
C_{\rm sca} = \frac{2}{\pi k_0}
\int_{0}^{2\pi}|f(\varphi)|^2\,d\varphi.
\label{eq:Csca_far}
\end{equation}

The extinction cross-section is obtained from the optical theorem,
which in the present convention gives\footnote{The right-hand side of Eq.~\eqref{eq:optical_theorem} has the sign opposite to that of the formula given in~\cite{Hulst2018}. This is because Ref.~\cite{Hulst2018} assumes the time dependence $\exp(\ii \omega t)$, whereas here we use $\exp(-\ii \omega t)$.}~\cite{Hulst2018}
\begin{equation}
C_{\rm ext}=
-\frac{4}{k_0}\,\Ree\,f(\varphi_0).
\label{eq:optical_theorem}
\end{equation}
The absorption cross-section is then found from
\begin{equation*}
C_{\rm abs}=C_{\rm ext}-C_{\rm sca}.
\end{equation*}

To use these relations in the present problem, we need the explicit expression for $f(\varphi)$. To this end, in Eqs.~\eqref{eq:ext_field} and \eqref{eq:psi_sca_p}, we express $\Hn{n}(k_0r_p)\ee^{\ii n\varphi_p}$ in terms of the coordinates of the observation point whose position vector in the global coordinate system is $\vect{r}$. Since, by definition, $k_0r_p \gg 1$ in the far field for any $p$, we may use the asymptotic expression~\eqref{eq:hankel_asymptotic} for $\Hn{n}(k_0r_p)$.

Next, we write $\vect{r_p}=\vect{r}-\vect{c}_p$, square this expression, and take the square root. This gives
\begin{eqnarray*}
 r_p  &= & r\left(1 -2\frac{\vect{r}\cdot \vect{c}_p }{r^2}+ \frac{c_p^2}{r^2}\right)^{1/2} \approx \\
   & & \approx  r - (x_p\cos\varphi + y_p\sin\varphi).
\end{eqnarray*}
Here $(x_p,y_p)$ are the Cartesian coordinates of the tip of $\vect{c}_p$ in the global coordinate system, and $\varphi$ is the polar angle of the observation point $\vect{r}$. We have expanded the square root in a Taylor series, keeping only terms linear in the small parameter $c_p/r$. Taking into account that, to the same accuracy, $\varphi_p \approx \varphi$ for any $p$, we obtain

\begin{equation}
f(\varphi)=
\sum_{p=1}^{P}\sum_{n=-\infty}^{\infty}
A_{np}\,\ee^{-\ii n\pi/2}
\ee^{-\ii k_0(x_p\cos\varphi + y_p\sin\varphi)}\,
\ee^{\ii n\varphi}.
\label{eq:scattering_amplitude}
\end{equation}
In our implementation of the optical theorem, the forward-scattering amplitude contains by definition the additional prefactor $\sqrt{2/(\pi k_0)}\,\ee^{-\ii\pi/4}$ that follows from the asymptotic form~\eqref{eq:far_field_amplitude}. Therefore, the expression
\[C_{\rm ext} = -\sqrt{8\pi/k_0}\,{\rm Re}\!\left[\ee^{\ii\pi/4}\psi^{\rm (sca,\,fwd)}_\infty\right]\]
and Eq.~\eqref{eq:optical_theorem} are identical.

\section{Interim summary\label{sec:summary}}

We now finish the analytical part and turn to the numerical methods used to solve the system of equations for the modal coefficients and to compute the fields and other physical characteristics of the problem. Before doing so, it is useful to summarize the analytical results.

We have obtained $P$ equivalent representations of the solution, one in each local cylindrical coordinate system associated with a cylinder of the cluster. In such a local coordinate system, the solution of the exterior boundary-value problem is given by Eqs.~\eqref{eq:psi_sca_p}, \eqref{eq:psi_ext}, and \eqref{eq:general_cyl_expantion_out}, where the coefficients~$B_{np}$ are expressed in terms of~$A_{np}$ and the incident-wave coefficients~$B_{np}^{(0)}$ according to Eq.~\eqref{eq:Bpn}. The coefficients $A_{np}$ themselves are found by solving the linear system~\eqref{eq:self_consistent}.

Although formally equivalent expressions for the external field are obtained for any choice of the local coordinate system, in actual computations it is convenient to choose the value of $p$ corresponding to the cylinder near which the field structure is of greatest interest. The interior boundary-value problem is described by Eqs.~\eqref{eq:int_field}, \eqref{eq:Mie_diagonal_D}, and \eqref{eq:t}, and is ultimately also expressed through the coefficients $A_{np}$.

When describing the overall field structure, it is convenient to return to the global coordinate system. This transition from any local system is achieved by a simple shift of the origin.

\section{Description of the numerical algorithm \label{sec:numerics}}

\subsection{Truncation of the series
\label{sec:truncation}}

The first step in actual applying the exact solution above is truncating the infinite multipole series at a finite order $N$. The physical basis for this truncation is the well-known fact that, for scattering of a monochromatic plane wave by a single cylinder with size parameter $x^{(0)}\equiv k_0 a$, the contribution of multipoles of order~$n$ to the scattered radiation decreases exponentially for $|n| \gg k_0 a$. At such values of $n$, the incident wave hardly excites the corresponding modes, and the scattering coefficients $|s_n|$ rapidly tend to zero. Therefore, in our numerical computations the index $n$ satisfies the condition $|n| \leq N$, i.e., $M = 2N+1$ harmonics are retained for each cylinder, with the same value of $N$ for all cylinders in the cluster.

The initial value of the truncation order was chosen using the heuristic formula
\begin{equation*}
{N_{\mathrm{start}} =
\left[
8 + x_{\max}
+ 5\,x_{\max}^{1/3}
\right];\;
x_{\max} \equiv k_0\max_p\left[ a_p|\sqrt{\eps_p}|\right],}
\end{equation*}
where square brackets denote taking the integer part.

The sufficiency of the chosen~$N$ was checked by convergence of the results: after increasing~$N$ by~2, the field at the probe points (see Sec.~\ref{sec:verification}) had to change by no more than a prescribed tolerance, typically about~$1\%$. When necessary, for example to resolve sharp spatial variations of the field, the accuracy was increased. In the specific nanotrimer computations presented below, the required accuracy was achieved at the fixed value $N=18$.

\subsection{Matrix formulation of the problem
\label{sec:matrix_form}}

After truncation, system~\eqref{eq:self_consistent} for a cluster of $P$ cylinders becomes a system of $P\cdot M$ linear equations with $P\cdot M$ unknowns, where $M = 2N+1$ coefficients $A_{np}$ are assigned to each cylinder. It is convenient to write this system in compact block form. For each cylinder~$p$, we introduce the column vector of its unknown scattering coefficients {$\vect{A}_p = (A_{-Np},\;\ldots,\;A_{Np})$} of length~$M$ and the analogous vector of incident-wave coefficients $\vect{B}_p^{(0)} = (B_{-Np}^{(0)},\;\ldots,\;B_{Np}^{(0)})$.

To avoid confusion with the Poynting vector, we denote the diagonal matrix\footnote{All matrices are denoted by boldface symbols.} of the scattering coefficients of the $p$-th cylinder by a Gothic letter: \mbox{$\bm{\mathfrak{S}}_p = \diag(s_{-Np},\;\ldots,\;s_{Np})$,} and the translation matrix with elements~\eqref{eq:T_pq} by $\mathbf{T}_{pq}$. These matrices have size $M\times M$. In the general case of an ensemble of $P$ cylinders, the system~\eqref{eq:self_consistent} becomes
\begin{equation}
\sum_{q=1}^P \bm{\mathfrak{A}}_{pq} \vect{A}_q = \bm{\mathfrak{S}}_p\,\vect{B}_p^{(0)}, \quad p = 1, \ldots, P,
\label{eq:block_system_general}
\end{equation}
where the blocks of the global system matrix $\bm{\mathfrak{A}}$ of size \mbox{$P\cdot M \times P\cdot M$} are defined as
\begin{equation*}
\bm{\mathfrak{A}}_{pq} =
\begin{cases}
\mathbf{I}, & p = q, \\
-\bm{\mathfrak{S}}_p \mathbf{T}_{pq}, & p \neq q,
\end{cases}
\end{equation*}
and $\mathbf{I}$ is the $M \times M$ identity matrix.

For the three-cylinder cluster considered below, $P=3$, and the block system can be written explicitly as
\begin{equation}
\underbrace{
\begin{pmatrix}
\mathbf{I} & -\bm{\mathfrak{S}}_1 \mathbf{T}_{12} & -\bm{\mathfrak{S}}_1 \mathbf{T}_{13} \\
-\bm{\mathfrak{S}}_2 \mathbf{T}_{21} & \mathbf{I} & -\bm{\mathfrak{S}}_2 \mathbf{T}_{23} \\
-\bm{\mathfrak{S}}_3 \mathbf{T}_{31} & -\bm{\mathfrak{S}}_3 \mathbf{T}_{32} & \mathbf{I}
\end{pmatrix}}_{\displaystyle\bm{\mathfrak{A}}}
\underbrace{\begin{pmatrix}
\vect{A}_1 \\ \vect{A}_2 \\ \vect{A}_3
\end{pmatrix}}_{\displaystyle\bm{\mathcal{A}}}
=
\underbrace{\begin{pmatrix}
\bm{\mathfrak{S}}_1\,\vect{B}_1^{(0)} \\
\bm{\mathfrak{S}}_2\,\vect{B}_2^{(0)} \\
\bm{\mathfrak{S}}_3\,\vect{B}_3^{(0)}
\end{pmatrix}}_{\displaystyle\bm{\mathcal{B}}^{(0)}}.
\label{eq:block_system}
\end{equation}
The identity diagonal blocks describe the self-response of each cylinder, while the off-diagonal blocks $-\bm{\mathfrak{S}}_p\mathbf{T}_{pq}$ describe their coupling through multiple scattering.

In actual computations, the matrix~$\bm{\mathfrak{A}}$ and the vector~$\bm{\mathcal{B}}^{(0)}$ are constructed explicitly. The off-diagonal blocks $-\bm{\mathfrak{S}}_p \mathbf{T}_{pq}$ for $p\neq q$ are formed as the product of the diagonal matrix $\bm{\mathfrak{S}}_p$ and the translation matrix $\mathbf{T}_{pq}$, whose elements $\Hn{m-n}(k_0 R_{pq})\,\ee^{\ii(m-n)\theta_{pq}}$ are precomputed once for each pair $(p,q)$ and stored in memory. The diagonal matrix $\bm{\mathfrak{S}}_p$ is stored not as a matrix but as a vector of length $M$ composed of the values $s_{np}$, $|n|\leq N$, computed from Eq.~\eqref{eq:s} using the recurrence relation for the derivatives of the Bessel functions; see Sec.~\ref{sec:bessel_deriv}. Representing $\bm{\mathfrak{S}}_p$ as a one-dimensional array rather than an $M\times M$ matrix is an implementation detail that eliminates unnecessary memory access in multiplication \mbox{by $\mathbf{T}_{pq}$.}

\subsection{Use of mirror symmetry\label{sec:symmetry}}

If the cluster geometry and the direction of the incident wave have mirror symmetry with respect to a plane passing through the $z$ axis, then the electromagnetic field has the same symmetry. This makes it possible to reduce the computational cost by determining the field configuration in only one of the mirror halves.

Let the $y$ axis of the global coordinate system coincide with the incident-wave vector. Then the mirror symmetry transformation corresponds to \mbox{$x \to -x$,} which in polar coordinates means $\varphi \to \pi - \varphi$. Since in each local coordinate system the $x$ and $y$ axes are parallel to the corresponding axes of the global system, this transformation induces the simultaneous transformation $\varphi_p \to \pi - \varphi_p$ in \emph{all} local systems.

Suppose that, among the $P$ cylinders of the cluster, the axes of $K$ cylinders lie in the mirror-symmetry plane, while the remaining $P-K$ do not. Then mirror reflection maps the $K$ cylinders onto themselves and splits the remaining cylinders into symmetric pairs. In this case, symmetry means that the transformation does not change the field $\psi$. Applying it to Eq.~\eqref{eq:psi_sca_p} for the $K$ cylinders whose axes lie in the symmetry plane, we obtain

\begin{eqnarray*}
   & & \sum_{n=-N}^N A_{np}\,\Hn{n}(k_0 r_p)\,\ee^{\ii n\varphi_p} =
\end{eqnarray*}
\begin{eqnarray*}
 & =&\sum_{m=-N}^N A_{mp}\,\Hn{m}(k_0 r_p)\,(-1)^m\,\ee^{-\ii m\varphi_p} = \\
 & =&\sum_{n=-N}^N A_{-np}\,\Hn{-n}(k_0 r_p)(-1)^{-n}\,\ee^{\ii n\varphi_p} = \\
   & =&\sum_{n=-N}^N A_{-np}\,\Hn{n}(k_0 r_p)\,\ee^{\ii n\varphi_p},
\end{eqnarray*}
where we relabeled the dummy index $m$ as $-n$ and used the identity $\Hn{-n}(z) = (-1)^n \Hn{n}(z)$. Orthogonality of the functions $\Hn{n}$ then gives $A_{np} = A_{-np}$. For symmetric pairs of cylinders, $p$ and $p'$, whose axes do not lie in the symmetry plane, an analogous relation holds: $A_{np} = A_{-np'}$.

This reduces the number of independent coefficients $A_{np}$ for each of the $K$ cylinders from $2N+1$ to $N+1$ and allows one to exclude $(P-K)/2$ cylinders lying on one side of the mirror-symmetry plane (recall that, by definition, $P-K$ is an even number). Thus, the total number of independent variables is reduced from $P(2N+1)$ to
\[
K(N+1) + (2N+1)\frac{P-K}{2} \equiv NP + \frac{K+P}{2}.
\]

In particular, for the trimer discussed below, see Sec.~\ref{sec:results}, $P=3$ and $K=1$, so the number of independent coefficients $A_{np}$ is reduced from $3(2N+1)$ to $3N+2$. This reduction was used in all computations presented below. The results were compared with the solution of the full system in Sec.~\ref{sec:verification}. Since the full and reduced systems are built by different code branches with independent matrix operations, agreement of their solutions at the level of $\sim 10^{-13}$ verifies the absence of programming errors in the assembly of the blocks $\bm{\mathfrak{S}}_p\mathbf{T}_{pq}$ and the correctness of the transfer of the coefficients of the mirror cylinder when forming the right-hand side of the equations.

\subsection{Condition number of the system matrix \label{sec:condition_number}}

The central indicator of the difficulty of solving system~\eqref{eq:block_system} numerically is its \emph{condition number}. For an arbitrary nonsingular complex matrix $\bm{\mathfrak{A}}$ of the size $n\times n$, it is defined as

\begin{equation*} \kappa(\bm{\mathfrak{A}}) = \norm{\bm{\mathfrak{A}}}\cdot\norm{\bm{\mathfrak{A}}^{-1}}, 
\end{equation*}
\noindent
where $\norm{\,\dots\,}$ is any consistent matrix norm. Unless stated otherwise, we use the spectral norm $\norm{\bm{\mathfrak{A}}}_2 = \sigma_{\max}$, so that
\begin{equation} \kappa(\bm{\mathfrak{A}}) = \frac{\sigma_{\max}(\bm{\mathfrak{A}})} {\sigma_{\min}(\bm{\mathfrak{A}})}, \label{eq:cond_svd}
\end{equation}
where $\sigma_{\max}$ and $\sigma_{\min}$ are the largest and smallest singular values of $\bm{\mathfrak{A}}$, i.e., the square roots of the largest and smallest eigenvalues of the Hermitian positive-semidefinite matrix $\bm{\mathfrak{A}}^\dagger\bm{\mathfrak{A}}$.

The meaning of $\kappa$ is as follows. If the right-hand side of the equations is specified with relative error $\delta$, then the relative error of the solution does not exceed $\kappa\delta$. In the present case, $\delta\sim u \approx 2.22\times 10^{-16}$, where $u$ is the machine epsilon of \texttt{float64} arithmetic. In terms of significant decimal digits, this means that in the computed coefficients $A_{np}$ no more than
\begin{equation*}
d_{\text{preserved}} \approx d_{\text{input}} - \log_{10}\kappa 
\end{equation*}
digits are preserved~\cite{higham2002accuracy}, where $d_{\text{input}}\approx 16$ for double precision. In particular, for \(\kappa\gtrsim 10^{16}\), one can expect, in the worst case, the loss of \emph{all} significant digits in the forward error of the solution. Therefore, a result obtained in \texttt{float64} arithmetic alone cannot be considered reliable without additional checks. This is precisely why the multilevel adaptive solver described in Sec.~\ref{sec:cascade} is needed.

\vspace*{5pt}
\subsubsection{Dependence on problem parameters}
\vspace*{5pt}

In the present problem, the condition number $\kappa(\bm{\mathfrak{A}})$ depends on three groups of parameters.

\emph{System size}. As the truncation order~$N$ of the multipole expansion increases, the number of modes \mbox{$M = 2N+1$} grows, and therefore so does the size of the matrix $\bm{\mathfrak{A}}$. The higher modes $|n|\gg k_0 a$ are weakly excited by the incident wave but are present in the matrix through the translation elements $(\mathbf{T}_{pq})_{nm}$, which contain high-order Hankel functions $\Hn{m-n}(k_0 R_{pq})$. At fixed argument $k_0 R_{pq}$, the modulus {$|\Hn{\ell}|$ grows factorially as $|\ell|\to\infty$, whereas $|J_\ell|$ decreases factorially}. This leads to an exponential spread in the moduli of the diagonal and off-diagonal matrix elements and to the corresponding growth of $\kappa$. Such growth has been confirmed numerically for the aluminum trimer discussed below: at $g/a=0.01$, where $g$ is the gap between cylinders, the values of $\kappa$ are $4.1\times10^{44}$ for $N=18$, $2.7\times10^{65}$ for $N=24$, and $2.5\times10^{87}$ for $N=30$. This corresponds to about $3.6$ orders of magnitude of growth in $\kappa$ per unit increase of $N$.

\emph{Cluster geometry}. For subwavelength cylinders and subwavelength gaps, the quantities $k_0 R_{pq}$ are small, at least for neighboring cylinders. In such cases, the functions $\Hn{\ell}$ behave as $(k_0 R_{pq})^{-|\ell|}$. This produces large off-diagonal elements of the translation matrix and ultimately determines the ill-conditioning of $\bm{\mathfrak{A}}$ for densely packed clusters of subwavelength cylinders. The conditioning deteriorates as neighboring cylinders approach each other.

\emph{Material parameters}. Near the resonances of individual cylinders, the denominators in Eq.~\eqref{eq:s} for the scattering coefficients $s_{np}$ become small, producing large elements in the diagonal blocks $\bm{\mathfrak{S}}_p$ and thus additionally increasing $\kappa$. For plasmonic materials with $\eps'<0$ and small $\eps''$, such as aluminum in the wavelength range discussed in Sec.~\ref{sec:results}, this effect is especially pronounced.

\subsubsection{Practical computation and use}

Direct computation of $\kappa$ from Eq.~\eqref{eq:cond_svd} by singular value decomposition requires $O(n^3)$ operations. For \mbox{$n = PM$} of order several hundred, this cost is not prohibitive. In well-conditioned and moderately conditioned regimes, the standard Python function \texttt{numpy.linalg.cond}~\cite{harris2020numpy} is used; it returns $\kappa$ in the spectral norm.

However, this method has a fundamental limitation. Singular value decomposition in \texttt{float64} arithmetic cannot provide reliable values of $\kappa$ if $\kappa  \gtrsim 10^{16}$: the computed smallest singular value $\sigma_{\min}$ cannot fall below the rounding-noise level $\sim u\,\sigma_{\max}$, where $u \approx 2.22\times 10^{-16}$ is the machine epsilon. As a result, the estimate of the ratio~\eqref{eq:cond_svd} saturates at $\sim 1/u \sim 10^{16}$ regardless of the true value of $\kappa$. For choosing the solver level, item~(i) below, this fact is usually sufficient: saturation near $10^{16}$ reliably indicates that the system is ill-conditioned and that extended-precision arithmetic is required. For quantitative determination of $\kappa$, however, \texttt{numpy.linalg.cond} is unsuitable in this regime. Therefore, the values of $\kappa$ in the range $10^{21}\ldots 10^{44}$ were obtained by recomputing the corresponding norms in the extended-precision arithmetic provided by \texttt{mpmath}~\cite{johansson2013mpmath}. The quantities obtained in this way agree with the analytical estimate $|\Hn{\ell}(k_0 R_{pq})|\sim(k_0 R_{pq})^{-|\ell|}$ for the growth of the off-diagonal translation-matrix elements, which provides an additional reliability check.

The value of $\kappa$ is used in two ways: (i)~to choose the solver level according to the threshold values described in Sec.~\ref{sec:cascade}; and (ii)~to adaptively determine the number of decimal digits $\mathtt{mp.dps}$ in the \texttt{mpmath} arithmetic using the heuristic formula
\begin{equation}
\mathtt{mp.dps} = \max\!\bigl(50,\; \lfloor \log_{10}\kappa \rfloor + 20\bigr),
\label{eq:dps_heuristic}
\end{equation}
which provides a margin of approximately $20$ significant decimal digits above the minimum needed to compensate for the loss of accuracy during factorization. In the ill-conditioned regime, the high-precision estimate of $\kappa$ must be substituted into Eq.~\eqref{eq:dps_heuristic}; using the saturated \texttt{float64}-SVD value would underestimate $\log_{10}\kappa$ and, consequently, the required number of working digits.

In applications to nano-optical phenomena, however, Eq.~\eqref{eq:dps_heuristic} is usually not needed. In particular, in the example discussed below, we use the fixed value \mbox{$\mathtt{mp.dps} = 120$,} which gives a margin of about $100$ decimal digits relative to double precision.

\vspace*{-6pt}
\subsubsection{Conditioning and symmetry reduction}
\vspace*{-3pt}

The transition from the full system of equations to the symmetry-reduced system described in Sec.~\ref{sec:symmetry} itself \emph{improves} the condition number of the matrix. This improvement results from projecting the original problem onto the symmetry subspace and eliminating the eigenvalues corresponding to antisymmetric modes, which do not contribute to the solution of a symmetric problem but do affect the matrix spectrum. Although the typical ratio $\kappa_{\text{full}} /\kappa_{\text{red}}$ is only a few units, the reduced system often remains within the applicability range of \texttt{float64} arithmetic at parameter values for which the full system already requires switching to \texttt{mpmath}.

\subsection{Residual and the criterion of solution quality\label{sec:residual}} 

After the vector $\bm{\mathcal{A}}$ is found, the \emph{relative residual} of system~\eqref{eq:block_system} is computed to monitor accuracy:

\begin{equation}
\rho = \max_{1 \leq p \leq P}\,
\frac{\displaystyle\max_{|n|\leq N}\bigl|
  \bigl(\bm{\mathfrak{A}}\bm{\mathcal{A}} - \bm{\mathcal{B}}^{(0)}\bigr)_{np}
  \bigr|}
{\displaystyle\max_{|n|\leq N}\bigl|\mathcal{B}^{(0)}_{np}\bigr|}.
\label{eq:residual}
\end{equation}
Here the numerator is the maximum absolute residual error over all equations of block~$p$, the denominator normalizes it to the amplitude of the right-hand side of the same block, and the outer maximum is taken over all $P$ cylinders. In the numerical solution, we use an adaptive four-level solver with successively increasing computational precision, described in detail in Sec.~\ref{sec:cascade}.

At each solver level, $\rho$ is computed. If $\rho$ exceeds the thresholds specified below, the solution is passed to the next level.

We emphasize, however, that a small value of $\rho$ indicates only a small \emph{backward error} in the linear-algebra step. It does not guarantee a small \emph{forward error} in the computed coefficients~$A_{np}$ or in the fields computed from them, because such errors may also arise from the finite truncation order $N$, approximate evaluation of the cylindrical functions, and summation of the series. The multistage adaptive solver only improves the accuracy of the solution of the linear system; the forward error is related to the accuracy of the coefficients of this system. These issues must be considered together, as done in the following sections.

\emph{Additional instability indicator}. If the maximum coefficient amplitude $\max_{np}|A_{np}|$ exceeds a prescribed threshold, typically $10^4$, the configuration is flagged as potentially unstable regardless of the $\rho$ value.
\vspace*{-4pt}
\subsection{Four-level solver structure\label{sec:cascade}}
\vspace*{-4pt}

To obtain a reliable solution over the entire parameter range, including configurations near resonances and/or with subwavelength gaps, a four-level procedure has been implemented. Each subsequent level is invoked only when the accuracy of the previous one is insufficient.

\paragraph{Level~1. Standard LU decomposition (\texttt{numpy.linalg.solve}).\footnote{\samepage{Here and below, the parentheses after each paragraph title indicate the specific Python function that directly implements the corresponding solver level.}}}
The system~\eqref{eq:block_system_general} is solved by LU factorization with partial pivoting in \texttt{float64} arithmetic (IEEE~754 standard, approximately 16 significant decimal digits). This is implemented by calling \texttt{numpy.linalg.solve}~\cite{harris2020numpy}, which internally uses the LAPACK routine \texttt{zgesv}~\cite{anderson1999lapack}. For well-conditioned configurations, far from resonances and with moderate gaps, level~1 gives a residual $\rho \lesssim 10^{-12}$, and no transition to higher levels is required.

\paragraph{Level~2. Matrix scaling and iterative refinement.}
If a preliminary estimate indicates ill-conditioning, $\kappa(\bm{\mathfrak{A}}) \gtrsim 10^{12}$ (which is still much smaller than the indicated above saturation level $10^{16}$), two-sided diagonal scaling (matrix equilibration) is applied to the matrix $\bm{\mathfrak{A}}$ and the right-hand side $\bm{\mathcal{B}}^{(0)}$:
\begin{equation*}
\bm{\mathfrak{\tilde{A}}} = \bm{G}_r\,\bm{\mathfrak{A}}\,\bm{G}_c,
\qquad
\bm{\mathcal{\tilde{B}}}^{(0)} = \bm{G}_r\,\bm{\mathcal{B}}^{(0)},
\end{equation*}
where the diagonal matrices $\bm{G}_r$ and $\bm{G}_c$ reduce the spread of the row and column norms of $\bm{\mathfrak{A}}$.

Here the \emph{row norm} of the $i$-th row of a matrix is the maximum modulus of its elements, $\max_j |A_{ij}|$, and the \emph{column norm} is defined analogously. The matrices \mbox{$\bm{G}_r = \operatorname{diag}(1/r_1,\ldots,1/r_{MP})$} and \mbox{$\bm{G}_c = \operatorname{diag}(1/c_1,\ldots,1/c_{MP})$} are constructed sequentially. First, $r_i = \max_j|A_{ij}|$, so that after row normalization $\max_j|A_{ij}/r_i| = 1$ for each row~$i$. Then, $c_j = \max_i|A_{ij}/r_i|$, which ensures \mbox{$\max_i\bigl|(\bm{\mathfrak{\tilde{A}}})_{ij}\bigr|=1$} for each column~$j$. Strict two-sided equality of all norms to unity is generally not achieved in one pass; when necessary, the procedure is repeated iteratively as two-sided diagonal equilibration~\cite{higham2002accuracy}.

The scaled system is solved by LU factorization, after which the obtained solution $\bm{\mathcal{\tilde{A}}}^{(0)}$ is refined by at most five residual-correction iterations. At each step, the correction and the updated approximation are computed as
\begin{equation*}
\bm{\delta}^{(k)} =
  \bm{\mathfrak{\tilde{A}}}^{-1}\!
  \left(\bm{\mathcal{\tilde{B}}}^{(0)}
      - \bm{\mathfrak{\tilde{A}}}\,\bm{\mathcal{\tilde{A}}}^{(k)}\right);\;
\bm{\mathcal{\tilde{A}}}^{(k+1)} =
  \bm{\mathcal{\tilde{A}}}^{(k)} + \bm{\delta}^{(k)},
\end{equation*}
where $\bm{\mathcal{\tilde{A}}}^{(k)}$ denotes the $k$-th approximation, \mbox{($k = 0, 1, \dots, 5$).}

After each step, the residual $\rho\!\left(\bm{\mathcal{\tilde{A}}}^{(k+1)}\right)$ is computed. If it does not decrease relative to $\rho\!\left(\bm{\mathcal{\tilde{A}}}^{(k)}\right)$, the iterations are terminated early. The approximation $\bm{\mathcal{\tilde{A}}}^{(k_{\rm min})}$ with the smallest achieved residual is accepted as the result. The original coefficients are recovered according to the relation $\bm{\mathcal{A}} = \bm{G}_c\,\bm{\mathcal{\tilde{A}}}^{(k_{\rm min})}$.

It is important to emphasize that, at this level, the residual is computed in standard \texttt{float64} arithmetic. Thus, this is standard iterative refinement~\cite{anderson1999lapack}, not a mixed-precision method. To obtain a substantial gain in a mixed-precision scheme, the residual must be computed in arithmetic of higher precision than that used for the factorization~\cite{higham2002accuracy}.

For ill-conditioned problems, level~2 may still be insufficient. In such cases, control is passed to level~3.

\paragraph{Level~3. Arbitrary-precision arithmetic (\texttt{mpmath}).}
For $\kappa(\bm{\mathfrak{A}}) > 10^{16}$, the entire matrix $\bm{\mathfrak{\tilde{A}}}$ and the right-hand side are recomputed from scratch using the Python library \texttt{mpmath}~\cite{johansson2013mpmath} with $\mathtt{mp.dps} = 120$ working decimal digits rather than converted from \texttt{float64}; this prevents the transfer of rounding errors from lower solver levels into the high-precision representation.

The solution obtained by \texttt{mpmath.lu\_solve} is converted back to \texttt{float64}, after which the residual is checked again. Level~3 increases the reliability of solutions of ill-conditioned systems: the intermediate computations are carried out with 120 decimal digits, suppressing the accumulation of rounding errors during factorization, while the final coefficients $A_{np}$ are rounded to standard double precision.

\paragraph{Level~4. Exact elimination over the field $\QQI$ (\texttt{sympy.DomainMatrix}).}
For the most critical configurations, such as subwavelength gaps near resonances, the code provides the possibility of an \emph{exact} solution of the system, completely eliminating rounding errors during matrix inversion. We now explain what is meant by this.

In modern numerical computations, real numbers are stored in memory in floating-point format (IEEE~754 standard). In this format, each number is represented as $(-1)^s\times m\times 2^e$, where $s\in\{0,1\}$ is the sign, $m$ is an integer specifying the significant digits (mantissa), and the integer $e$ determines the magnitude, playing, in base~2, the same role as the power of 10 in the familiar notation $1.23\times 10^{n}$. Standard double precision (\texttt{float64}) allocates 53 binary digits to the mantissa, corresponding to about 16 significant decimal digits, and 11 digits to $e$, with $-1022 \leq e \leq 1023$, which covers magnitudes from about $10^{-308}$ to $10^{308}$. Arithmetic operations on such numbers are accompanied by rounding to 53 binary digits. In ill-conditioned linear systems, these small errors may accumulate and eventually lead to a catastrophic loss of accuracy.

The key idea of this solver level is to eliminate such rounding by using the fact that any IEEE~754 number has an {\it exact\/} representation as a fraction of the form $m/2^e$, with an integer numerator and a power of two in the denominator, i.e., as a {\it rational number}. Complex numbers with rational real and imaginary parts form the \emph{field of Gaussian rational numbers} $\QQI$, i.e., numbers of the form $a + \ii b$, where $a$ and $b$ are rational. This set is closed under addition, subtraction, multiplication, and division: the sum, difference, product, and quotient of two Gaussian rational numbers are again Gaussian rational numbers.

Thus, the real and imaginary parts of all elements of the equilibrated matrix $\bm{\mathfrak{\tilde{A}}}$ and right-hand side  vector $\bm{\mathcal{\tilde{B}}}^{(0)}$ can be stored as rational numbers and represented {\it exactly\/} as ordinary fractions.
Therefore, Gaussian elimination applied to the system $\bm{\mathfrak{\tilde{A}}}\,\bm{\mathcal{\tilde{A}}} = \bm{\mathcal{\tilde{B}}}^{(0)}$ with elements from $\QQI$ is performed \emph{exactly}: all intermediate quantities remain rational fractions, and no rounding errors occur during elimination~\cite{meurer2017sympy}.

In practice, this is implemented in Python using \texttt{sympy.DomainMatrix}~\cite{meurer2017sympy}. The matrix $\bm{\mathfrak{\tilde{A}}}$ is converted to the field $\QQI$ by command \texttt{.convert\_to(QQ\_I)}, after which \texttt{.solve(b)} is called with the exact right-hand side vector. The exact solution is then converted back to \texttt{float64}, and the final residual~\eqref{eq:residual} is computed.

The level~4 implementation was verified independently: fraction-free Bareiss elimination over $\mathbb{Z}[\ii]$ at $g/a=0.01$, $N=8$ (dimension 51) gave a \emph{strictly zero} residual \mbox{$\bm{\mathfrak{A}}\bm{\mathcal{A}}-\bm{\mathcal{B}}^{(0)}=0$} and agreement with level~3 with an error $\|x_4-x_3\|\sim10^{-121}$ at $\mathtt{mp.dps}=120$.

At the same time, one must distinguish clearly between what exact elimination ensures and what it does not. Exact elimination \emph{completely eliminates} the backward error of the matrix-inversion step: if the elements of $\bm{\mathfrak{\tilde{A}}}$ and $\bm{\mathcal{\tilde{B}}}^{(0)}$ are specified exactly, the solution $\bm{\mathcal{\tilde{A}}}$ is found without rounding errors during elimination. However, exact elimination \emph{does not remove} the forward error caused by errors in the input data. For example, if the matrix elements contain special-function evaluation errors of order {$u\sim 10^{-16}$}, the machine epsilon of \texttt{float64} arithmetic, then exact elimination finds the exact solution of an \emph{inexactly} formulated system~\cite{higham2002accuracy}.

Thus, the rigorous statement about level~4 is as follows: it minimizes the backward error of the linear-algebra step, but it does not eliminate other unavoidable sources of error, namely (i)~the error due to truncation of the multipole series, which is controlled by the convergence criterion described in Sec.~\ref{sec:truncation}; (ii)~the initial error in the computation of special functions at the machine-epsilon level, which is present in the matrix elements and cannot be corrected by any postprocessing; and (iii)~the error of summing the series when computing the field with the obtained modal coefficients. For configurations in which the accuracy of special-function computation is especially important, for example at \mbox{$|x_p^{(c)}| \gg 1$}, the elements of $\bm{\mathfrak{S}}_p$ and $\mathbf{T}_{pq}$ can be recomputed using \texttt{mpmath} before invoking level~4, which provides a fully high-precision computational chain.

In principle, level~4 is redundant because the same computations can also be performed at level~3. The difficulty is that the accuracy of level~3 depends on the prescribed value of $\mathtt{mp.dps}$, and the heuristic estimates used for this value do not guarantee that it is sufficient. Level~4 is free from this drawback. Its main purpose is therefore the cross-validation of level~3.

\section{Computation of the electromagnetic field and of the Poynting vector\label{sec:poynting}}
\subsection{Computation of the field in different regions}

After the coefficients $\{\vect{A}_p\}$ have been found, the electromagnetic field at an arbitrary point $\vect{r}$ is computed as follows. If the point is outside all cylinders, the total external-field representation~\eqref{eq:psi_1} is used, with the scattered contribution summed over all $P$ cylinders according to Eqs.~\eqref{eq:ext_field} and \eqref{eq:psi_sca_p}. If the point is inside cylinder~$p$, the Bessel-function series~\eqref{eq:int_field} is used, with the coefficients $\vect{D}_p$ computed by Eq.~\eqref{eq:Mie_diagonal_D}.

To prevent numerical artifacts near the cylinder boundaries, three different procedures are used in the code depending on the purpose of the computation. In \emph{visualizations} of the scalar field on a regular grid, a ring of half-width $bw \cdot a_p$ is introduced between the interior region of the cylinder, where the field is described by Bessel functions, and the exterior region, where the field is a superposition of the incident plane wave and of the scattered partial waves described by Hankel functions. Here $bw \in [0.04,\,0.10]$. In this ring, the field and its spatial derivatives are obtained as a linear combination of the interior and exterior representations and therefore do not satisfy Maxwell's equations. This region is used \emph{only for rendering images} and is not used for quantitative diagnostics.

In quantitative computations of the Poynting vector in the exterior region, a buffer \mbox{$\delta_{\rm ext} = 10^{-3}\,\min_p a_p$} is applied. At the stage of local refinement of the maximum on the surface, $10^{-1}\,\delta_{\rm ext}=10^{-4}\,\min_p a_p$ is used. For the grid map of $|\vect{S}|$, the buffer $\delta_{\rm grid} = 5\cdot 10^{-3}\,\min_p a_p$ is used. All numerical values of $|\vect{S}|$ presented in this work were obtained without using the interpolation ring.

\subsection{Computation of special functions and their derivatives
\label{sec:bessel_deriv}}

The Bessel functions $J_n(z)$ and Hankel functions $H_n^{(1)}(z)$ for integer $n$ are computed over the whole array of orders \mbox{$-(N+1)\leq n \leq N+1$} simultaneously using the standard Python functions \texttt{scipy.special.jv} and \texttt{scipy.special.hankel1}. The coefficients $s_{np}$ and $t_{np}$ are cached using \texttt{functools.lru\_cache}. This avoids repeated solution of the single-cylinder problem. The special functions themselves are recomputed for each set of probe points, but in vectorized form over the index~$n$, which provides the main performance gain compared with element-by-element computation. The derivatives are computed using the recurrence relation
\begin{equation*}
F_n'(z) = \tfrac{1}{2}\bigl[F_{n-1}(z) - F_{n+1}(z)\bigr],
\end{equation*}

\noindent which requires the values $F_{n\pm 1}$ in advance. This is why $J_n(z)$ and $H_n^{(1)}(z)$ are tabulated over the range $[-N-1, N+1]$ rather than $[-N, N]$. This approach provides accuracy of order machine epsilon, unlike finite-difference approximation of the derivative, which would introduce an additional error $\mathcal{O}(h^2)$.

\subsection{Vectorized computation of the field at probes
and on the visualization grid}

The computation of the field $\psi(\vect{r})$ is vectorized simultaneously over observation points and the modal index $n$. For each cylinder~$p$, the contribution of the scattered field at all probe points can be written as a matrix product:

\begin{eqnarray}
    \psi^{(\rm sca)}_{pl}
& = &\sum_{n=-N}^{N} A_{np}\,\Hn{n}(k_0 r_{pl})\,\ee^{\ii n\varphi_{pl}} \equiv \label{eq:field_vectorized} \\
    & \equiv & \bigl(\vect{A}_p\bigr)^T \bigl[\vect{h}_{pl} \odot \vect{e}_{pl}\bigr],\nonumber
\end{eqnarray}

\noindent where $l$ is the observation-point index, $\vect{h}_{pl}$ and $\vect{e}_{pl}$ are vectors of length~$M$ containing $\Hn{n}(k_0 r_{pl})$ and $\ee^{\ii n\varphi_{pl}}$, respectively, and $\odot$ denotes elementwise multiplication. The matrices $[\vect{h}_{pl}]$ and $[\vect{e}_{pl}]$ have size $M \times N_{\rm probe}$, where $N_{\rm probe}$ is the number of probes. These expressions are precomputed once before the optimization iterations and stored in RAM.

The computation of the field~\eqref{eq:field_vectorized} and its spatial derivatives, which determine the Poynting vector~$\vect S$, on the visualization grid is organized blockwise. The grid is partitioned into fragments of moderate size, $160\times160$ nodes in the computations presented here, and the formula is applied to each fragment in vectorized form. This allows each block to be computed independently. The partitioning is purely technical; its only purpose is to keep memory consumption constant regardless of the total number of nodes, making it possible to construct maps of the Poynting vector and its streamlines at arbitrarily high resolution.

The spatial derivatives of $\psi$ are computed analytically from the same cylindrical-function series. For a field of the form

\begin{equation}
\psi(r,\varphi) = \sum_{n=-N}^{N} c_n\,F_n(\varkappa r)\,\ee^{\ii n\varphi},
\end{equation}
where $F_n = \Jn{n}$ inside the cylinder and $F_n=\Hn{n}$ outside, and where \mbox{$\varkappa = k_p$} or $k_0$, respectively, the radial and azimuthal derivatives are

\begin{equation*}
\frac{\partial\psi}{\partial r}
= \varkappa\sum_n c_n\,F_n'(\varkappa r)\,\ee^{\ii n\varphi}, \quad
\frac{\partial\psi}{\partial\varphi}
= \ii\sum_n c_n\,n\,F_n(\varkappa r)\,\ee^{\ii n\varphi}.
\end{equation*}
The Cartesian derivatives are obtained from these derivatives by the standard relations
\begin{equation}
\frac{\partial\psi}{\partial x}
= \cos\varphi\,\frac{\partial\psi}{\partial r}
  - \frac{\sin\varphi}{r}\,\frac{\partial\psi}{\partial\varphi};\;
\frac{\partial\psi}{\partial y}
= \sin\varphi\,\frac{\partial\psi}{\partial r}
  + \frac{\cos\varphi}{r}\,\frac{\partial\psi}{\partial\varphi}.
\label{eq:cart_grads}
\end{equation}

Equations~\eqref{eq:cart_grads} contain division by $r$ and are formally singular at $r=0$, i.e., on the cylinder axis. This singularity is only apparent. From the expansion $\psi = \sum_n c_n J_n(\varkappa r)e^{in\varphi}$ and the asymptotics ${|J_n(\varkappa r)|} \approx (\varkappa r/2)^{|n|}/|n|!$ and $J_0(\varkappa r) \approx 1 + \mathcal{O}\left((\varkappa r)^2\right)$ as $\varkappa r\to 0$, it follows that the derivatives on the right-hand sides of Eq.~\eqref{eq:cart_grads} are determined by the $J_{\pm 1}$ terms. All other harmonics contribute only higher-order terms that vanish as $\varkappa r\to 0$. Thus, we obtain

\begin{eqnarray*}
   & & \left.\frac{\partial\psi}{\partial r}\right|_{r=0}
= \frac{\varkappa}{2}\bigl(c_1\,\ee^{\ii\varphi} - c_{-1}\,\ee^{-\ii\varphi}\bigr), \\
   & & \left.\frac{1}{r}\frac{\partial\psi}{\partial\varphi}\right|_{r=0}
= \frac{\ii\varkappa}{2}\bigl(c_1\,\ee^{\ii\varphi} + c_{-1}\,\ee^{-\ii\varphi}\bigr).
\end{eqnarray*}
Substituting these expressions into Eq.~\eqref{eq:cart_grads} cancels the dependence on $\varphi$, yielding
\[
\left.\frac{\partial\psi}{\partial x}\right|_{r=0}
= \frac{\varkappa}{2}(c_1 - c_{-1}),
\quad
\left.\frac{\partial\psi}{\partial y}\right|_{r=0}
= \frac{\ii\varkappa}{2}(c_1 + c_{-1}),
\]
which confirms the regularity of these derivatives on the cylinder axis.

\subsection{Streamlines of the Poynting vector field\label{sec:streamlines}}
The field lines of the Poynting vector field $\vect{S}$, which by analogy with hydrodynamics we call \emph{streamlines}, are defined as curves whose tangent at every point has the same direction as $\vect{S}$. Let $\vect{r}_{_{\vect{S}}}(t)$ be the parametric equation of such a curve, where $t$ is a scalar parameter. The tangency condition requires that the vectors $d\vect{r}_{_{\vect{S}}}/dt$ and $\vect{S}$ be collinear, giving $d\vect{r}_{_{\vect{S}}}/dt = \mathrm{const}\cdot\vect{S}$.

This equation is inconvenient for numerical integration. At singular points of the field, streamlines may intersect, which is possible only if $\vect{S}=0$. Near such points, $S \equiv |\vect{S}|$ is small, and direct numerical integration of the unnormalized equation leads to excessively small steps. Normalizing the right-hand side by $S$ removes this problem without changing the direction of the vector field. Choosing the scale of the parameter $t$ so that the proportionality coefficient between $d\vect{r}_{_{\vect{S}}}/dt$ and $\vect{S}/S$ is unity, we obtain
\begin{equation}
\frac{d\vect{r}_{_{\vect{S}}}}{dt}
  = \frac{\vect{S}\!\left(\vect{r}_{_{\vect{S}}}\right)}{S}.
\label{eq:streamline_ode}
\end{equation}
Since the right-hand side is a unit vector, the parameter $t$ coincides with the arc length along the streamline. The integration is carried out in both directions from the initial point corresponding to $t=0$, i.e., for both $t>0$ and $t<0$, and the two tracks are joined to form a single continuous streamline.

\paragraph{Integration method.} %
The Cauchy problem~\eqref{eq:streamline_ode} is solved numerically by an explicit Runge--Kutta method with adaptive step-size selection based on the embedded Bogacki--Shampine pair RK3(2)~\cite{bogacki1989pair}. This method belongs to the class of methods with automatic error control by embedded formulas of different orders. Its Butcher tableau is
\begin{equation*}
\renewcommand{\arraystretch}{1.2}
\begin{array}{c|cccc}
0     &       &       &      & \\
1/2   & 1/2   &       &      & \\
3/4   & 0     & 3/4   &      & \\
1     & 2/9   & 1/3   & 4/9  & \\[3pt]\hline\rule{0pt}{12pt}
b^{(3)}   & 2/9   & 1/3   & 4/9  & 0   \\
b^{(2)}   & 7/24  & 1/4   & 1/3  & 1/8
\end{array}
\end{equation*}
where $b^{(3)}_i$ and $b^{(2)}_i$ are the weights of the third- and second-order quadrature formulas, respectively. The method has the FSAL (First Same As Last) property: the fourth stage of the current step is identical to the first stage of the next one, reducing the computational cost to three evaluations of the right-hand side of Eq.~\eqref{eq:streamline_ode} per integration step.

Let $\vect{r}_{m+1}^{(3)}$ and $\vect{r}_{m+1}^{(2)}$ be the approximations to the exact solution $\vect{r}(t_m + h_m)$ obtained with the weights $b^{(3)}$ and $b^{(2)}$, respectively. The higher-order approximation $\vect{r}_{m+1} = \vect{r}_{m+1}^{(3)}$ is taken as the advanced solution, i.e., local extrapolation is used~\cite{hairer1993solving}. The local error estimate is constructed as the difference between the two approximations:
\begin{equation}
\vect{e}_{m+1} = \vect{r}_{m+1}^{(3)} - \vect{r}_{m+1}^{(2)}.
\label{eq:bs23_err}
\end{equation}

Accuracy is controlled using the normalized error, which includes both absolute ($\varepsilon_{\rm abs}$) and relative ($\varepsilon_{\rm rel}$) tolerances:
\begin{eqnarray*}
   & &\hspace*{-4mm} E_{\rm norm} =\\ 
   & & \hspace*{-4mm}= \left\{
  \sum_{\alpha=x,y}
  \left(\frac{e_\alpha}
       {\varepsilon_{\rm abs}
        + \varepsilon_{\rm rel}\,
          \max\!\bigl(|r_\alpha^{(m)}|,\,|r_\alpha^{(m+1)}|,\,10^{-6}\bigr)}
  \right)^{\!2}
\right\}^{1/2}\!,\nonumber
\end{eqnarray*}
where $e_\alpha$ are the components of the vector $\vect{e}_{m+1}$ in Eq.~\eqref{eq:bs23_err}. The tolerances are $\varepsilon_{\rm abs} = 10^{-5}$ and \mbox{$\varepsilon_{\rm rel} = 5\times 10^{-4}$.} The threshold $10^{-6}$ in the denominator prevents division by zero.

\emph{Adaptive step-size control.}
If $E_{\rm norm} \leq 1$, the step is accepted, the point $\vect{r}_{m+1}$ is added to the history, and the new step is chosen as
\begin{equation}
h_{m+1} = \min\Bigl(h_{\max},\;
h_m \cdot \min\left(5,\; 0.9\,E_{\rm norm}^{{-1/3}}\right)\Bigr).
\label{eq:step_accept}
\end{equation}
For $E_{\rm norm} < 10^{-12}$, i.e., for a practically machine-zero error, the factor is set to~$2.5$. If $E_{\rm norm} > 1$, the required accuracy has not been reached, and the step is repeated with the reduced value
\begin{equation}
h_{\rm retry} = \max\Bigl(h_{\min},\;
h_m \cdot \max\left(0.1,\; 0.9\,E_{\rm norm}^{{-1/3}}\right)\Bigr).
\label{eq:step_reject}
\end{equation}
The step is also accepted unconditionally if the current value of $h$ is already equal to $h_{\min}$, which guarantees advancement even in regions with sharp field variations. After twenty sequential unsuccessful attempts, a forced Euler step is performed:
$\vect{r}_{m+1} = \vect{r}_m + h_{\min}\,\vect{S}(\vect{r}_m)/|\vect{S}(\vect{r}_m)|$, after which the adaptive integration is resumed with the initial step~$h_0$.

The exponent $-1/3$ in Eqs.~\eqref{eq:step_accept} and \eqref{eq:step_reject} is determined by the order of the error estimate of the embedded pair. The local error~\eqref{eq:bs23_err} is of order $O(h^{\hat{p}+1})$, where $\hat{p}=2$ is the order of the lower-order solution. When the step is changed by a factor $\sigma$, the error scales as $\sigma^{\hat{p}+1}$. Hence the condition $E_{\rm norm}\,\sigma^{\hat{p}+1} = 1$ gives the optimal factor $\sigma_{\rm opt} = E_{\rm norm}^{-1/(\hat{p}+1)} = E_{\rm norm}^{-1/3}$~\cite{hairer1993solving}. The same exponent is used both for increasing the step after a successful step and for reducing it after an unsuccessful one, ensuring a uniform adaptation strategy. The factor $0.9<1$ is a safety coefficient that helps avoid immediate rejection after step-size increase.

\begin{table*}[htb]
\centering
\caption{Parameters of the test configuration of the aluminum-nanocylinder trimer.}
\label{tab:params}
\begin{tabular}{@{}ll@{}}
\toprule
\hline
\rule[10pt]{0pt}{0pt}
Parameter & Value \\[1pt]
\midrule
Wavelength $\lambda$ & 116~nm \\
Polarization & $H_z$ \\
Radii of the cylinders $a_1 = a_2 = a_3 {\equiv a}$ & {10.0~nm} \\
Minimum gap between cylinders $g$ & 5.0~nm \\
Gap-to-radius ratio $g/a$ & {0.5} \\
Size parameter $k_0 a$ & {0.542} \\
Side length of the triangle $d = 2a + g$ & {25.0~nm} \\
Permittivity of Al ($\lambda = 116$~nm)
  & $\eps \approx  -0.974 + 0.086\ii$ \\
Truncation order \(N\) (adaptive) & 18 \\
Number of equations for the coefficients $A_{np}$  & $3(2\times 18 +1)=111$  \\
Number of equations for the coefficients $A_{np}$ in the symmetry-reduced system \hspace*{6pt} & $3\times 18+2 = 56$ \\
Number of points in the visualization grid & ${400}\times{400}$ \\
\bottomrule
\end{tabular}
\end{table*}

\paragraph{Stopping conditions.}
The integration stops when one of the following events occurs:
\begin{itemize}
  \item[(i)] the trajectory leaves the computational domain $[-L,L]^2$;
  \item[(ii)] $|\vect{S}|^2 < 10^{-60}$, corresponding to the vicinity of a topological singular point, i.e., a practical zero \mbox{of $\vect{S}$;} 
  \item[(iii)] stagnation occurs, meaning that over the last 20 steps the displacement is less than $5\%$ of the arc length of the traversed segment, which is typical near a singular point. 
\end{itemize}

No dead zone is introduced near the interface. The integration is carried out in a through-boundary-crossing mode, in which $\vect{S}$ is computed analytically on both sides of the boundary, with the normal components smoothly matched on the boundary itself. The tangential component of $\vect{S}$ may be discontinuous at the boundary, which plays an important role in the formation of the Poynting-vector-field topology.

To conclude the discussion of the numerical part of the method, we note that despite the validation and cross-validation techniques described above, the final evidence of convergence and high computational accuracy is the satisfaction of physical tests. Section~\ref{sec:verification} is devoted to this analysis.

\vspace*{3pt}
\section{Aluminum trimer}\label{sec:results}
\vspace*{2pt}
\subsection{Configuration parameters}
\vspace*{2pt}

As noted in the Introduction, to demonstrate the capabilities of the method we consider a trimer of three identical aluminum nanocylinders (nanowires) of circular cross-section whose axes are located at the vertices of an equilateral triangle lying in the plane perpendicular to these axes. The origin of the global coordinate system is chosen at the centroid of the triangle. The vertex coordinates are

\begin{equation}
\vect{c}_p = R\bigl(\cos\theta_p,\,\sin\theta_p\bigr),\quad
\theta_p = \frac{\pi}{2} + \frac{2\pi(p-1)}{3},
\label{eq:centers}
\end{equation}
where $ p = 1,\;2,\;3$. The incidence angle is $\varphi_0=\pi/2$, so the incident wave vector $\vect{k}_0$ is parallel to the $y$ axis, and $R = d/\sqrt{3} = (2a+g)/\sqrt{3}$. The computational parameters are given in Table~\ref{tab:params}. 

The permittivity of aluminum at $\lambda = 116$~nm, where $\lambda \equiv 2\pi/k_0$, was taken from Palik's tabulated data~\cite{palik1998} and set equal to \mbox{$\eps \approx -0.974 + 0.086\,\ii$,} corresponding to the plasmonic scattering regime ($\eps' < 0$) with moderate dissipative losses ($\eps'' \ll 1$).

The magnitudes of the scattering coefficients for the first few values of $n$ at the chosen permittivity and size parameter are presented in Table~\ref{tab:sn}. The magnitude of $|s_1|$ is much larger than the others, meaning that the selected parameters correspond to the vicinity of the dipole resonance of a single cylinder. The table also shows the rapid decrease of $|s_n|$ for $n\geq 3$.

\begin{table}[!ht]
\centering
\vspace*{-6pt}
\caption{Magnitudes of the scattering coefficients $|s_n|$ for the values of $\eps$ and $k_0a$ given in Table~\ref{tab:params}.}
\label{tab:sn}
\begin{tabular}{l|lllll}
\toprule
\hline
\rule[10pt]{0pt}{0pt}
Parameter & \multicolumn{5}{c}{Value}\\ [1pt]
\hline
$n$      & 0      & 1      & 2      & 3                    & 4                    \\
$|s_n|$   & 0.014 & 0.730 & 0.095 & $1.82\cdot 10^{-3}$ & $1.225\cdot 10^{-5}$ \\
\bottomrule
\end{tabular}
\end{table}

We emphasize that the parameters were chosen only to demonstrate the capabilities of the method clearly. The results presented here do not claim to provide a complete description of the various near-field structures arising in this problem, nor do they optimize the geometry for record enhancement of the fields $\vect{E}$, $\vect{H}$, or $\vect{S}$. These questions will be discussed in a separate publication.
\paragraph{Choice of initial points.}
To visualize the topology of the Poynting vector field as completely as possible, the initial integration points are taken from several independent sources. In the exterior region, these are: (a)~a uniform rectangular grid, with $3 \times 8$ nodes in the half-plane $x \geq 0$ when mirror symmetry is used (the result is then reflected), or $5 \times 8$ nodes over the whole domain $[-L,L]^2$ otherwise; (b)~16 points on 6 concentric rings around each cylinder at the radii {$r/a \in \{1.005,\; 1.01,\; 1.02,\;1.04,\; 1.08,\; 1.15\}$}; and (c)~$5 \times 5$ points in each intercylinder gap, corresponding to 5 longitudinal sections and 5 transverse offsets within $\pm 0.3a$.

For streamlines inside the cylinders, a grid of {$4 \times 10$} points in the radial {$r/a \in \{0.3,\; 0.5,\; 0.7,\; 0.9\}$} and angular ($\varphi_n=2\pi n/10,$ \mbox{$n \in \{1,\;2,\ldots,10\}$}) variables is used inside each cylinder.

Initial points closer than {$1.003\,a$} to the surface of a neighboring cylinder are excluded to avoid immediate penetration into that cylinder. When mirror symmetry is used, seeds with $x<0$ are excluded, and the obtained streamlines are reflected into the left half-plane. The final set is thinned out: for each streamline, its median point, corresponding to the middle of the arc length, is computed; if the median points of two lines are separated by less than $0.03a$, one of them is removed.

The initial steps depend on the seed type. In the case considered here, $h_0 = 0.03\,a$ for the far field, $h_0 = 0.02\,a$ for the near-surface region and gaps, and $h_0 = 0.015\,a$ for the cylinder interiors. We use $h_{\max} = 6\,h_0$ and \mbox{$h_{\min} = 10^{-5}\,a$.}

\subsection{Near-field map and streamlines}

\begin{figure*}
  \centering
  \includegraphics[width=.9\textwidth]{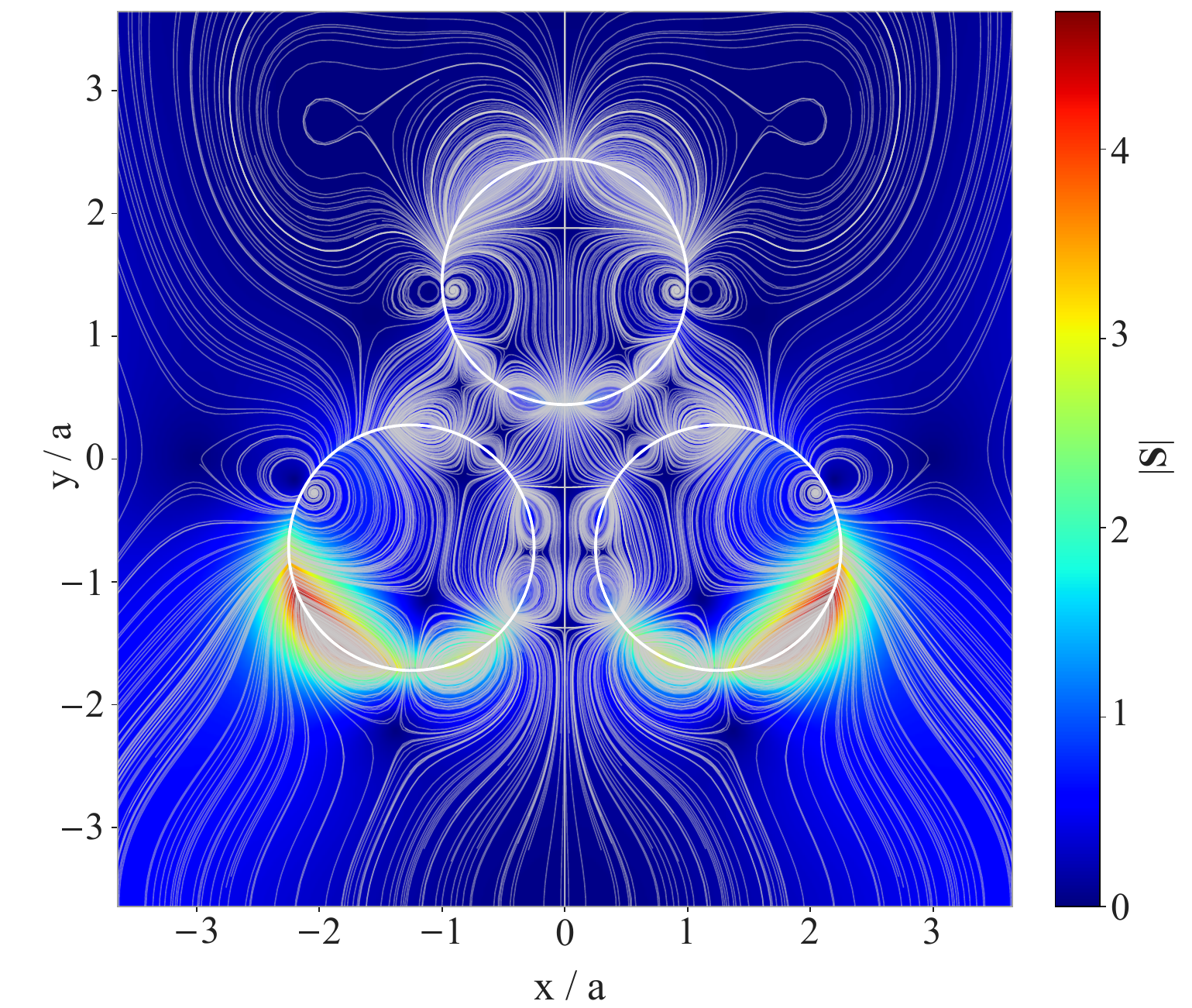}
  \caption{Map of the Poynting vector field for scattering of a linearly polarized monochromatic plane wave by the trimer at the parameter values given in Table~\ref{tab:params}. The cylinder surfaces are indicated by white circles. The incident wave propagates along the $y$ axis, from bottom to top in the figure. The vector $\vect{E}$ lies in the plane of the figure \mbox{($H_z$~polarization)}. The color scale shows the magnitude of the Poynting vector normalized to its magnitude in the incident wave.}
  \vspace*{-8pt}
  \label{fig:Poynting}
\end{figure*}

Figure~\ref{fig:Poynting} shows a map of the Poynting vector field near the trimer with streamlines superimposed. The topology of the energy-flux density is complex. In the far exterior region, the streamlines generally follow the propagation direction of the incident wave (from bottom to top in the figure). Near the cluster, however, they are strongly deformed, with energy concentrated in surface plasmons around which characteristic vortex structures are formed.

Two important properties of this field map are fairly general and are also observed for other parameter values, provided the mirror symmetry with respect to the $y$ axis and the orientation of the trimer relative to the incident radiation are preserved. First, a branch of the separatrix runs along the $y$ axis, separating the basins of attraction of the field lines associated with the right and left cylinders. As a result, no energy exchange occurs between these two cylinders. By contrast, the energy exchange between each of them and the upper cylinder is quite strong. Second, the cylinders at the base of the trimer, i.e., the upstream cylinders, screen the upper cylinder. Therefore, the characteristic values of the Poynting-vector magnitude inside and near the upper cylinder are smaller than those for the base cylinders. However, the situation may change in the vicinity of cooperative mode resonant frequencies---a topic not discussed here. 

\vspace*{5pt}
\subsection{Far field. Scattering indicatrix}
\vspace*{5pt}

An important far-field characteristic of the scattered radiation describing its angular dependence is the scattering indicatrix. To remove the trivial radial dependence caused by the divergence of the scattered cylindrical wave, the indicatrix should be normalized. We normalize it to the forward-scattering intensity, which in the global coordinate system used here corresponds to \mbox{$\varphi = 90^\circ$}.

In the example considered below, the scattering indicatrix was computed in several independent ways. The results were compared to check the accuracy and stability of the computations.

First, the indicatrix is related to the scattering amplitude by the expression
\begin{equation}\label{eq:I_via_f}
  I(\varphi) = \frac{|f(\varphi)|^2}{|f(\varphi_0)|^2}.
\end{equation}

\begin{figure}[!ht]
  \centering
  \includegraphics[width=.8\columnwidth]{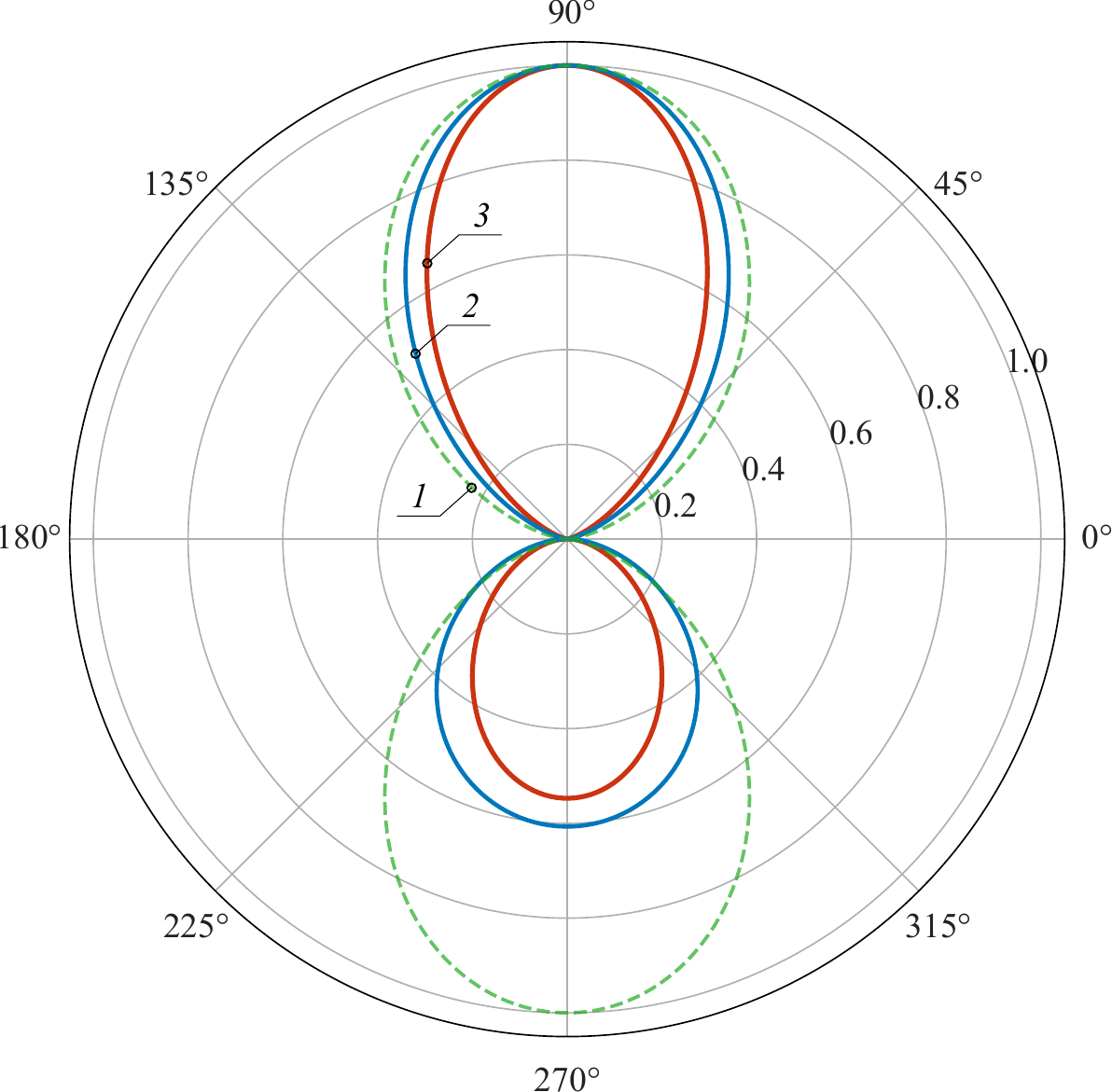}
  \caption{Scattering indicatrices normalized to the forward-scattering intensity. Curve 1 corresponds to electric-dipole radiation, curve 2 to a single cylinder, and curve 3 to the trimer. In cases 2 and 3, the parameter values are those given in Table~\ref{tab:params}.}
  \label{fig:indicatrix}
\end{figure}

Second, the indicatrix can be computed from first principles. A circle of radius $100\lambda$ centered at the origin of the global coordinate system was drawn. The incident field was subtracted from the total electromagnetic field, leaving only the scattered one. For this field, the radial component of the Poynting vector was computed as a function of the observation angle $\varphi$. The circle radius was then doubled, and the computations were repeated.

The curves obtained by these methods were compared. The comparison showed that all three curves agree with high accuracy\footnote{For the reader's convenience, a detailed discussion of the comparison criteria is given in Sec.~\ref{sec:verification}.}.

The scattering indicatrix obtained in this way is shown in Fig.~\ref{fig:indicatrix}. For comparison, the scattering indicatrices of a single cylinder and of an electric dipole are also shown. The latter is described by the well-known expression \mbox{$I=\sin^2\varphi$} in the coordinate system used here.

The trimer indicatrix is close to that of a single cylinder, and both are close to the dipole-radiation pattern. Thus, despite the complex near-field structure shown in Fig.~\ref{fig:Poynting}, the dipole mode makes the main contribution to the far-field directivity pattern of the trimer. This is consistent with the dominant role of the dipole scattering coefficient in the multipole expansion at the selected parameter values; see Table~\ref{tab:sn}.

\subsection{Scattering, absorption, and extinction cross-sections}

While the indicatrix determines the angular distribution of the scattered radiation, the integral characteristics of the phenomenon are the scattering, absorption, and extinction cross-sections $C_{\rm sca}$, $C_{\rm abs}$, and $C_{\rm ext}$. It is convenient to introduce the dimensionless efficiencies $Q_{\rm ext,\,sca,\,abs}$ by dividing the corresponding cross-sections by the geometric cross-section of the scattering object. In the two-dimensional geometry considered here, for the trimer this normalization corresponds to division by three cylinder diameters:
\begin{equation*}
   Q^{(3)}_{\rm ext,\, sca,\,abs} \equiv \frac{C^{(3)}_{\rm ext,\, sca,\,abs}}{6a},
\end{equation*}
where the superscript $(3)$ denotes the trimer.

For comparison, we also use the efficiencies of a single cylinder:
\begin{equation*}
  Q^{(1)}_{\rm ext,\, sca,\,abs} \equiv \frac{C_{\rm ext,\, sca,\,abs}^{(1)}}{2a}.
\end{equation*}
The individual absorption cross-sections of the cylinders forming the trimer are also of interest. They are denoted by $C^{(3,p)}_{\rm abs}$, where $p$ is defined according to Eq.~\eqref{eq:centers}; thus, $p=1$ corresponds to the apex of the trimer, while $p=2,3$ correspond to the base cylinders. Owing to symmetry, $C^{(3,2)}_{\rm abs} = C^{(3,3)}_{\rm abs}$. Since we compare the dissipation in each trimer cylinder with that in a single cylinder, it is convenient to define
\begin{equation*}
  Q^{(3,p)}_{\rm abs} \equiv \frac{C_{\rm abs}^{(3,p)}}{2a}.
\end{equation*}
Here $C^{(3)}_{\rm abs} = \sum_{p} C^{(3,p)}_{\rm abs}$, which gives \mbox{$\sum_{p=1}^{3} Q^{(3,p)}_{\rm abs} = 3\,Q^{(3)}_{\rm abs}$;} the factor 3 appears because $Q^{(3)}_{\rm abs}$ and $Q^{(3,p)}_{\rm abs}$ use different normalizations.

As in the computation of the indicatrix, when several equivalent expressions for the cross-sections were available, all of them were used. The values obtained by different methods were compared to monitor the quality of the computations; see Sec.~\ref{sec:verification}.

The results are presented in Table~\ref{tab:cross_sections}. All trimer efficiencies are noticeably smaller than the corresponding quantities for a single cylinder. This is because, for the selected parameters, the base cylinders screen the apex cylinder from the incident radiation. This screening is partially compensated by energy fluxes from the lower cylinders to the upper one, as seen in Fig.~\ref{fig:Poynting}. The compensation, however, is incomplete, while the energy transferred from the lower cylinders to the upper one reduces both the electromagnetic dissipation in the lower cylinders and the efficiency of their scattering.

\begin{table}[!ht]
\centering
\caption{Comparison of the scattering, absorption, and extinction efficiencies of the trimer and of a single cylinder at the parameters of Table~\ref{tab:params}
{($\lambda=116$~nm, $a=10$~nm, \mbox{$g=5$~nm}; aluminum, $H_z$~polarization).}
The trimer efficiencies $Q^{(3)}$ are normalized to the total transverse size of the three cylinders, $6a$; the single-cylinder efficiencies $Q^{(1)}$ and per-cylinder efficiencies $Q^{(3,p)}$ are normalized to $2a$.}
\label{tab:cross_sections}
{\setlength{\tabcolsep}{2pt}
\begin{tabular}{lcc}
\toprule
\hline
\rule[10pt]{0pt}{0pt}
 & Single cylinder & Trimer \\[2pt]
\hline
$Q_{\rm ext}$       & 5.217 & 2.767 \\
$Q_{\rm sca}$       & 4.018 & 2.237 \\
$Q_{\rm abs}$       & 1.199 & 0.532 \\
\hline
$Q^{(3,1)}_{\rm abs}$ & --- & 0.182 \\
$Q^{(3,2)}_{\rm abs}$ & --- & 0.708 \\
$Q^{(3,3)}_{\rm abs}$ & --- & 0.708 \\
\bottomrule
\end{tabular}}
\end{table}

It is also interesting to trace how the scattering characteristics of the trimer approach those of a single cylinder as the distance between the cylinders increases. The corresponding results are presented in Table~\ref{tab:cross_sections_2}.

For a single cylinder, exact expressions for the cross-sections in terms of the scattering coefficients $s_n$ are available~\cite{bohren1983,Hulst2018}:
\begin{equation}
Q_{\rm sca}^{(1)} = \frac{2}{k_0 a}\sum_{n}|s_n|^2,\qquad
Q_{\rm ext}^{(1)} = -\frac{2}{k_0 a}\,\Ree\sum_{n} s_n,
\label{eq:single_cyl_cs}
\end{equation}
together with the expression for $C_{\rm abs}^{(1)}$ following from these formulas\footnote{The coefficient $s_n$ introduced here has the sign opposite to that of the scattering coefficient traditionally used in such problems; this explains the minus sign in the expression for $Q_{\rm ext}^{(1)}$ in Eq.~\eqref{eq:single_cyl_cs}.} and from the definition of the extinction cross-section. Comparison of the values computed by the proposed method with $P=1$ and those obtained from Eq.~\eqref{eq:single_cyl_cs} gave zero machine error, which also confirms the high accuracy of the computations.

A reliable substantiation of the accuracy of the proposed algorithm, however, requires more systematic verification. We now turn to this issue.

\begin{table}[h]
\centering
\caption{Scattering, absorption, and extinction efficiencies of the equilateral trimer at different gap values $g$ ($g = d - 2a$), together with the corresponding single-cylinder values. Here $Q = C / \sigma_{\rm geom}$, where $\sigma_{\rm geom} = 2a$ for a single cylinder and $\sigma_{\rm geom} = 6a$ for the trimer. Parameters: $\lambda = 116$~nm, $a = 10$~nm, $\varepsilon = -0.974 + 0.086\,\ii$, and \mbox{$H_z$~polarization.} The truncation order is $N = 18$; the arithmetic precision was selected automatically by the four-stage adaptive solver. The column $Q_{\rm ext}/Q_{\rm ext}^{(1)}$ gives the ratio of the trimer extinction efficiency to that of a single cylinder and therefore measures cooperative screening.}
\label{tab:cross_sections_2}
{\setlength{\tabcolsep}{6pt}
\begin{tabular}{@{}rrcccc@{}}
\toprule
\hline
\rule[16pt]{0pt}{0pt}
$g$, nm & $g/a$
  & $Q_{\rm ext}$ & $Q_{\rm sca}$ & $Q_{\rm abs}$
  & $\dfrac{Q_{\rm ext}}{Q_{\rm ext}^{(1)}}$ \\[12pt]
\hline
\multicolumn{6}{@{}l}{\itshape Trimer} \\[2pt]
 5 & 0.50 & 2.767 & 2.237 & 0.532 & 0.531 \\
10 & 1.00 & 2.977 & 2.419 & 0.559 & 0.571 \\
20 & 2.00 & 3.420 & 2.768 & 0.652 & 0.656 \\
50 & 5.00 & 4.031 & 3.203 & 0.828 & 0.773 \\
\hline
\multicolumn{6}{@{}l}{\itshape Single cylinder} \\[2pt]
--- & --- & 5.217 & 4.018 & 1.199 & --- \\
\bottomrule
\end{tabular}}
\end{table}

\subsection{Comprehensive accuracy verification\label{sec:verification}}

The accuracy of the method was checked using a series of independent tests---each verifying a separate aspect of the computational scheme. Some of these tests have already been mentioned above. For convenience, we collect all of them here.

\paragraph{Convergence with respect to the truncation order~$N$.}
The dependence of $\max|H_z|$ and $\max|\vect{S}|$ on the truncation order~$N$ shows the emergence of a plateau at $N \gtrsim 18$. Increasing $N$ from the working value $18$ to $26$ changes the values of $|\vect{S}(\vect{r}_{\rm max})|$ at the fixed field-maximum point $\vect{r}_{\rm max}$ by no more than $\sim 10^{-3}$ on the absolute scale, corresponding to a relative accuracy $\delta|\vect{S}(\vect{r}_{\rm max})|/|\vect{S}(\vect{r}_{\rm max})|$ of order $10^{-4}$. The fast convergence at $N>18$  is due to the subwavelength size parameter $k_0 a \approx 0.54$: for $N \gg k_0 a$, the Bessel functions $J_n(k_0 a)$ decrease factorially, and the higher multipoles are exponentially suppressed.

However, besides the size parameter, the required value of $N$ depends strongly on the minimum gap $g$ between the cylinders. For large gaps, $g \gg a$, the interaction between the cylinders is suppressed, the higher multipoles are practically not excited, and moderate values of $N$ are sufficient. Conversely, as the cylinders approach each other ($g\ll a$) the field in the gap gets an increasingly sharp spatial profile, and high multipoles must be included to reproduce it.

Mathematically, this is related to the fact that the translation-matrix elements $T^{(pq)}_{nm}$ in Eq.~\eqref{eq:T_pq} contain the Hankel functions $\Hn{m-n}(k_0 R_{pq})$, whose modulus grows as the argument $k_0 R_{pq} = k_0(2a + g)$ decreases. This strengthens the coupling between higher modes of neighboring cylinders and slows the convergence of the multipole expansion as $g$ decreases. It explains why, for $g \ll a$, many multipoles are required to describe the field in the gap,  notwithstanding the small size parameter.\footnote{In the limit of touching cylinders, $g=0$, a singularity arises at the point of contact. In such a situation, even convergence of the multipole expansion itself becomes a separate question. This case requires a special analysis and is not considered here; see also the remark in Sec.~\ref{sec:geometry}.}

Thus, as $g$ decreases, $N$ must be increased to achieve the prescribed accuracy. As a result, the condition number $\kappa(\bm{\mathfrak{A}})$ of the matrix in system~\eqref{eq:block_system}, defined by Eq.~\eqref{eq:cond_svd}, grows. With traditional numerical methods for solving the linear system~\eqref{eq:block_system}, this leads to increased forward error and complicates the computation of fields in the regions most relevant for applications, where both the fields and their spatial derivatives may be large.

In our approach, this difficulty is overcome by reducing the backward error of the linear-algebra step and by using extended-precision arithmetic. This makes it possible to perform computations with the required accuracy for arbitrary cylinder configurations, including very small but finite gaps.

To illustrate this, Table~\ref{tab:NvsGap} gives the values of $\kappa$ at the working truncation order \mbox{$N=18$}, for $g/a \gtrsim 0.5$ and $k_0 a \approx 0.54$. For smaller gaps, \mbox{$g/a \lesssim 0.1$}, a larger $N$ may be required to resolve the sharp spatial variations of the field. The values of $\kappa$ given for such gaps should therefore be regarded as lower bounds corresponding to the specified truncation level.

\begin{table}[h]
\centering
\caption{Condition number $\kappa(\bm{\mathfrak{A}})$ of the matrix of system~\eqref{eq:block_system} at fixed $N = 18$ for different gap values $g$ ($\lambda = 116$~nm, $a = 10$~nm, and \mbox{$H_z$~polarization).} The column ``Solver level'' indicates the minimum level theoretically required to achieve \mbox{$\rho \lesssim 10^{-10}$.} In the demonstration considered below, $g/a = 0.5$, level~3 was used.}
\label{tab:NvsGap}
{\setlength{\tabcolsep}{12pt}
\begin{tabular}{@{}rrcc@{}}
\toprule
\hline
\rule[10pt]{0pt}{0pt}
$g/a$    & $g$, nm  & $\kappa(\bm{\mathfrak{A}})$ & Solver level \\[1pt]
\hline
\rule[10pt]{0pt}{0pt}
100.0   & 1000.0     & $5 \times 10^{1}$    & 1  \\
  5.00   &   50.0     & $3 \times 10^{21}$   & 3  \\
  2.00   &   20.0     & $1 \times 10^{31}$   & 3  \\
  1.00   &   10.0     & $3 \times 10^{36}$   & 3  \\
  0.50  &    5.0     & $2 \times 10^{40}$   & 3  \\
  0.10  &    1.0     & $1 \times 10^{44}$   & 3--4  \\
  0.01  &  0.1   & $4 \times 10^{44}$   & 3--4  \\
\bottomrule
\end{tabular}}
\end{table}

The last column indicates the minimum solver level, see Sec.~\ref{sec:cascade}, at which the residual $\rho \lesssim 10^{-10}$ is achieved. At $g/a = 100$, the system matrix is practically diagonal, $\kappa \approx 50$, and standard LU factorization at level~1 gives a residual at the machine-epsilon level.

As the gap decreases to $g/a = 0.5$, the condition number grows to about $10^{40}$. In this case, a reliable solution is obtained after switching to the third solver level, i.e., to arbitrary-precision \texttt{mpmath} arithmetic. At $g/a \lesssim 0.1$, the loss of significant digits during factorization makes arbitrary-precision arithmetic or exact elimination over $\QQI$ practically unavoidable. Note also that at $g/a \lesssim 0.01$, the growth of $\kappa$ with decreasing gap slows down because the contribution to conditioning from finite $N$ begins to dominate over the contribution from intercylinder coupling.

It is important that a large value of $\kappa$ alone does not imply an equally large error in the physical observables. The condition number refers to the full modal system, including weakly excited high harmonics whose contribution to the resulting field is small. Therefore, the final assessment of accuracy is based not on $\kappa$ alone but on the full set of physical and numerical tests described below.

\paragraph{Residual of the boundary conditions.}
The relative error in the boundary conditions~\eqref{eq:bc_psi} was computed on the surface of each cylinder at 360 equally spaced points. At each point, the quantities on the two sides of the equalities in Eq.~\eqref{eq:bc_psi} were computed independently from the exterior representation~\eqref{eq:psi_ext} and the interior representation~\eqref{eq:int_field}. Then, the two values were compared. The median and maximum relative errors over all comparisons were $3.6\times 10^{-6}$ and $2.4\times 10^{-5}$, respectively.
\paragraph{Scattering indicatrix.} 
For a quantitative comparison of the indicatrices obtained by the three methods described above, we computed
\begin{equation*}
  \Delta_{ij} = \max_{\varphi}\,
  \bigl|I_i(\varphi) - I_j(\varphi)\bigr|,
\end{equation*}
\noindent
where $I_i(\varphi)$ is the indicatrix normalized to its maximum, i.e., to the forward-scattering intensity, computed by the $i$-th method. Here $i=1$ corresponds to Eq.~\eqref{eq:I_via_f}, while $i=2,3$ correspond to the radial component of the Poynting vector at $r=100\lambda$ and $r=200\lambda$, respectively. The computation gave $\Delta_{23}\approx 10^{-3}$ and $\Delta_{12}\approx 2\times10^{-3}$, with $\Delta_{12}\approx 2\,\Delta_{13}$, in agreement with the expected decrease of the corrections\footnote{This decrease is related to the fact that, at finite $r$, the scattered wave is not purely transverse. For $\psi = \psi(r,\varphi)$, its gradient has a nonzero azimuthal component, which contains an extra factor of $1/(k_0r)$ relative to the radial component. Asymptotically, as $r\to\infty$, the wave becomes purely transverse. This azimuthal component generates a nonzero tangential component of $\vect{S}$, explaining the finite-radius corrections to the indicatrix in methods 2 and 3. Since $S_\varphi$ contains the same extra factor of $1/(k_0r)$ relative to $S_r$, the corresponding corrections decrease accordingly.} proportional to $1/(k_0r)$. Thus, the three independent methods of computing the indicatrix give results differing by no more than $0.2\%$.
\paragraph{Cross-sections.}
The scattering cross-section $C^{(3)}_{\rm sca}$ was determined in three independent ways: (i)~from the scattering amplitude $f(\varphi)$ using Eq.~\eqref{eq:Csca_far}; (ii)~by integrating the radial component of the Poynting vector of the scattered field over a circle of radius $R_1 = 100\lambda$; and (iii)~by the analogous integration over a circle of radius $R_2 = 200\lambda$.

Methods (i)--(iii) are not completely independent algebraically: all rely on the same expansion of the scattered field in cylindrical functions with coefficients $A_{np}$. They differ, however, in the summation procedure. In method~(i), the closed-form expression~\eqref{eq:scattering_amplitude} for the scattering amplitude is used, followed by analytical integration of its squared modulus. In methods~(ii) and~(iii), the total scattered field is computed pointwise on a finite radius circle by summing the series~\eqref{eq:ext_field} and \eqref{eq:psi_sca_p}, followed by numerical integration. The relative error for all three methods was about $3 \times 10^{-7}$. This test confirms the correct implementation of the phase factors in Eq.~\eqref{eq:Csca_far} and the absence of programming errors in the field-computation procedure at large $r$.

The values of $C^{(3,p)}_{\rm abs}$ were computed separately for each cylinder from the flux of the Poynting vector of the total external field through its lateral surface. The external field was computed using Eq.~\eqref{eq:psi_1}.

The extinction cross-section $C_{\rm ext}$ was computed in two independent ways. First, it was computed directly as
\begin{equation*}
C^{(3)}_{\rm ext}=C^{(3)}_{\rm sca}+C^{(3)}_{\rm abs}.
\end{equation*}
Second, it was computed from the two-dimensional optical theorem, Eq.~\eqref{eq:optical_theorem}. Comparison of the two values gave a relative error of about $10^{-6}$, providing another check of computational accuracy.

Methods (ii) and (iii) are the same as those used in the computation of the scattering indicatrix. However, for the indicatrix their discrepancy is about $10^{-3}$. It is three orders of magnitude larger than for $C_{\rm sca}$. This is because $C_{\rm sca}$ is an integral characteristic in which $S_r^{(\rm sca)}$ is effectively averaged over the angle $\varphi$, whereas the indicatrix $I(\varphi)$ is sensitive to the local intensity at each fixed angle.

\paragraph{Energy balance.}
We also checked the energy-flux balance through a closed circular contour of radius $R$ enclosing the whole cluster. The integral of the total-field flux over such a contour must be equal to $-C_{\rm abs}$:
\begin{equation*}
\oint_{\Gamma}\vect{S}_{\rm tot}\cdot\hat{\vect{n}}\,dl
= -\,C_{\rm abs},
\end{equation*}
where $\vect{S}_{\rm tot}$ is the Poynting vector of the total, incident plus scattered, field, and $\hat{\vect{n}}$ is the outward normal to the contour $\Gamma$. The relative error in this balance was about $4 \times 10^{-6}$.

\paragraph{Mirror symmetry.}
The equilateral trimer under incidence along the symmetry axis has mirror symmetry with respect to the plane \mbox{$x = 0$}. The solver uses this symmetry to reduce the number of independent coefficients $A_{np}$ from $3(2N+1)$ to $3N+2$, i.e., it reduces the number of equations in the corresponding linear system by $3N+1$; see Sec.~\ref{sec:symmetry}. An independent check showed that the scattering coefficients $A_{np}$ obtained from the full system agree with those obtained from the reduced system with a relative error not exceeding $5 \times 10^{-13}$, and that the values of the fields $\psi$ and $|\vect{S}|$ at the probe points are reproduced with the same accuracy.

The verification results for the configuration of Table~\ref{tab:params}, with $g=a/2$ and $N=18$, are summarized in Table~\ref{tab:verification}.
\begin{table}
\vspace*{-6pt}
\centering
\caption{Summary of verification results for \mbox{$H_z$~polarization,} $\lambda = 116$~nm, $g = a/2$, $N = 18$, solver level~3, and \mbox{$\mathtt{mp.dps} = 120$.}}
\label{tab:verification}
\begin{tabular}{@{}ll@{}}
\toprule
\hline
\rule[8pt]{0pt}{0pt}
Test & \hspace*{-15mm} Relative error \\[1pt]
\hline
Boundary conditions \\
\hspace*{12pt} (i) median error         &\hspace*{24pt} $3.6 \times 10^{-6}$ \\
\hspace*{12pt} (ii) maximum error     &\hspace*{24pt} $ 2.4 \times 10^{-5}$ \\
Scattering indicatrix                            &\hspace*{24pt} $ \sim 10^{-3}$ \\
Optical theorem                               &\hspace*{24pt} $7.5 \times 10^{-7}$ \\
Energy-flux balance                            &\hspace*{24pt} $3.9 \times 10^{-6}$ \\
Mirror symmetry\\ (in the coefficients $A_{np}$)                            &\hspace*{24pt} $\leq 5 \times 10^{-13}$ \\
\bottomrule
\end{tabular}
\end{table}

\section{Conclusion\label{sec:conclusion}}

In this work, we have proposed a complete semi-analytical scheme for computing light scattering by a cluster of an arbitrary number of infinitely long, parallel, homogeneous, non-overlapping and non-touching right circular cylinders. Each cylinder has an arbitrary complex permittivity $\eps_p$, constant within the cylinder, where $p$ is the cylinder index.

The main results are as follows:
\begin{enumerate}
\item Based on the exact analytical solution of the problem, we have obtained the compact block system~\eqref{eq:block_system_general} of linear equations for the cylindrical-function expansion coefficients. The system consists of the diagonal matrices of scattering coefficients $\bm{\mathfrak{S}}_p$ and the translation matrices $\mathbf{T}_{pq}$. It is valid for a cluster with an arbitrary number $P$ of cylinders.
\item We have presented a four-level adaptive solution strategy with control of conditioning and working precision:
LU decomposition $\to$ equilibration with iterative refinement $\to$ arbitrary-precision \texttt{mpmath} arithmetic $\to$ exact elimination over $\QQI$. This strategy provides numerically verified solutions over a broad parameter range, including configurations near resonances and/or with subwavelength cylinder sizes and gaps.
\item The method has been comprehensively verified using the example of scattering of a normally incident, $H_z$ linearly polarized monochromatic plane wave by a symmetric trimer of aluminum nanocylinders, with $a = 10$~nm, $\lambda = 116$~nm, $g = 5$~nm. The verification included convergence with respect to the truncation order $N$, control of the boundary-condition residual, analysis of the directivity pattern, and several physical tests.
\end{enumerate}

The method is computationally efficient, analytically controllable, and can be applied to arrays of $P$ cylinders of arbitrary configuration, oligomer chains, and ordered or disordered metamaterial lattices. Its semi-analytical character makes it promising for inverse design in nanophotonics~\cite{molesky2018}, fast preliminary screening of nano\-cluster geometries in SERS applications~\cite{hao2004,camden2008}, fluorescence enhancement, nanoplasmonics~\cite{knight2014,gerard2015}, and many other problems.

The approach developed here has several important advantages over the finite-element method (FEM)~\cite{jin2015,monk2003} and the finite-difference time-domain method (FDTD)~\mbox{\cite{taflove2005,yee1966}.} In particular:
\begin{itemize}
  \item[(i)] The Sommerfeld radiation condition is satisfied exactly through the asymptotics of the Hankel functions.
  \item[(ii)] The field at arbitrary spatial points is computed analytically without spatial interpolation or extrapolation.
  \item[(iii)] The near and far fields are computed from the same set of modal coefficients.
  \item[(iv)] There is no spatial grid, and the only discretization parameter is the truncation order $N$ of the multipole series.
  \item[(v)] The boundary conditions are satisfied analytically on the exact circular surfaces, eliminating the ``staircase'' or staircasing errors of FDTD grids.
  \item[(vi)] For narrow gaps between cylinders, accuracy is ensured by increasing $N$ and using the appropriate solver level, rather than by grid refinement.
 \end{itemize}

At the same time, the main drawback of the method is that it is applicable only to a strictly limited class of problems. Therefore, it does not replace universal methods such as \emph{FEM}, \emph{FDTD}, or \emph{BEM}, which can handle considerably more complex geometries and spatially inhomogeneous materials. Rather, it is a powerful complementary tool that is useful for parametric studies and for producing benchmark solutions for validating more universal numerical schemes.

We also point out that, after truncation of the multipole expansion, the size of the linear system for the modal coefficients is $PM$, where $P$ is the number of cylinders, $M=2N+1$, and $N$ is the truncation order. In this case, the computational cost of factorization scales as $\mathcal{O}((PM)^3)$, and memory consumption scales as $\mathcal{O}((PM)^2)$. This version of the algorithm is therefore aimed primarily at finite clusters of moderate size. As $P$ increases and/or the gaps between cylinders decrease, computational-cost limitations arise.

These limitations are not too restrictive in practice. For example, in the symmetric aluminum-trimer case considered here, confident convergence on a personal computer was achieved for a gap equal to $0.1\%$ of the cylinder radius.

Finally, the results were obtained for a cluster located in vacuum, but they are straightforwardly transferred to cylinders embedded in any homogeneous non-absorbing medium. Although the specific example of the aluminum trimer belongs to nano-optics, the domain of applicability of the developed approach is much broader. Within the convergence limits of the numerical solution, it is bounded only by the applicability of classical continuum electrodynamics and extends from the hard ultraviolet to the far radio band.

This work was carried out within the framework of the state assignment of Lomonosov Moscow State University. The numerical study was supported by the Russian Science Foundation (Grant No.~23-72-00037).

\bibliography{Tern_En}

\begin{thebibliography}{40}%
\makeatletter
\providecommand \@ifxundefined [1]{%
 \@ifx{#1\undefined}
}%
\providecommand \@ifnum [1]{%
 \ifnum #1\expandafter \@firstoftwo
 \else \expandafter \@secondoftwo
 \fi
}%
\providecommand \@ifx [1]{%
 \ifx #1\expandafter \@firstoftwo
 \else \expandafter \@secondoftwo
 \fi
}%
\providecommand \natexlab [1]{#1}%
\providecommand \enquote  [1]{``#1''}%
\providecommand \bibnamefont  [1]{#1}%
\providecommand \bibfnamefont [1]{#1}%
\providecommand \citenamefont [1]{#1}%
\providecommand \href@noop [0]{\@secondoftwo}%
\providecommand \href [0]{\begingroup \@sanitize@url \@href}%
\providecommand \@href[1]{\@@startlink{#1}\@@href}%
\providecommand \@@href[1]{\endgroup#1\@@endlink}%
\providecommand \@sanitize@url [0]{\catcode `\\12\catcode `\$12\catcode
  `\&12\catcode `\#12\catcode `\^12\catcode `\_12\catcode `\%12\relax}%
\providecommand \@@startlink[1]{}%
\providecommand \@@endlink[0]{}%
\providecommand \url  [0]{\begingroup\@sanitize@url \@url }%
\providecommand \@url [1]{\endgroup\@href {#1}{\urlprefix }}%
\providecommand \urlprefix  [0]{URL }%
\providecommand \Eprint [0]{\href }%
\providecommand \doibase [0]{https://doi.org/}%
\providecommand \selectlanguage [0]{\@gobble}%
\providecommand \bibinfo  [0]{\@secondoftwo}%
\providecommand \bibfield  [0]{\@secondoftwo}%
\providecommand \translation [1]{[#1]}%
\providecommand \BibitemOpen [0]{}%
\providecommand \bibitemStop [0]{}%
\providecommand \bibitemNoStop [0]{.\EOS\space}%
\providecommand \EOS [0]{\spacefactor3000\relax}%
\providecommand \BibitemShut  [1]{\csname bibitem#1\endcsname}%
\let\auto@bib@innerbib\@empty
\bibitem [{\citenamefont {Bohren}\ and\ \citenamefont
  {Huffman}(1983)}]{bohren1983}%
  \BibitemOpen
  \bibfield  {author} {\bibinfo {author} {\bibfnamefont {C.~F.}\ \bibnamefont
  {Bohren}}\ and\ \bibinfo {author} {\bibfnamefont {D.~R.}\ \bibnamefont
  {Huffman}},\ }\href {https://doi.org/10.1002/9783527618156} {\emph {\bibinfo
  {title} {Absorption and Scattering of Light by Small Particles}}}\ (\bibinfo
  {publisher} {Wiley},\ \bibinfo {address} {New York},\ \bibinfo {year}
  {1983})\BibitemShut {NoStop}%
\bibitem [{\citenamefont {Tsang}\ \emph {et~al.}(2000)\citenamefont {Tsang},
  \citenamefont {Kong},\ and\ \citenamefont {Ding}}]{tsang2000}%
  \BibitemOpen
  \bibfield  {author} {\bibinfo {author} {\bibfnamefont {L.}~\bibnamefont
  {Tsang}}, \bibinfo {author} {\bibfnamefont {J.~A.}\ \bibnamefont {Kong}},\
  and\ \bibinfo {author} {\bibfnamefont {K.-H.}\ \bibnamefont {Ding}},\ }\href
  {https://doi.org/10.1002/0471224286} {\emph {\bibinfo {title} {Scattering of
  Electromagnetic Waves: Theories and Applications}}}\ (\bibinfo  {publisher}
  {Wiley},\ \bibinfo {address} {New York},\ \bibinfo {year} {2000})\BibitemShut
  {NoStop}%
\bibitem [{\citenamefont {Rahmani}\ \emph {et~al.}(2004)\citenamefont
  {Rahmani}, \citenamefont {Chaumet},\ and\ \citenamefont
  {Bryant}}]{rahmani2004}%
  \BibitemOpen
  \bibfield  {author} {\bibinfo {author} {\bibfnamefont {A.}~\bibnamefont
  {Rahmani}}, \bibinfo {author} {\bibfnamefont {P.~C.}\ \bibnamefont
  {Chaumet}},\ and\ \bibinfo {author} {\bibfnamefont {G.~W.}\ \bibnamefont
  {Bryant}},\ }\href {https://doi.org/10.1086/383609} {\bibfield  {journal}
  {\bibinfo  {journal} {Astrophys. J.}\ }\textbf {\bibinfo {volume} {607}},\
  \bibinfo {pages} {873} (\bibinfo {year} {2004})}\BibitemShut {NoStop}%
\bibitem [{\citenamefont {Twersky}(1952)}]{twersky1952}%
  \BibitemOpen
  \bibfield  {author} {\bibinfo {author} {\bibfnamefont {V.}~\bibnamefont
  {Twersky}},\ }\href {https://doi.org/10.1121/1.1906845} {\bibfield  {journal}
  {\bibinfo  {journal} {J. Acoust. Soc. Am.}\ }\textbf {\bibinfo {volume}
  {24}},\ \bibinfo {pages} {42} (\bibinfo {year} {1952})}\BibitemShut {NoStop}%
\bibitem [{\citenamefont {Linton}\ and\ \citenamefont
  {Martin}(2005)}]{linton2005}%
  \BibitemOpen
  \bibfield  {author} {\bibinfo {author} {\bibfnamefont {C.~M.}\ \bibnamefont
  {Linton}}\ and\ \bibinfo {author} {\bibfnamefont {P.~A.}\ \bibnamefont
  {Martin}},\ }\href {https://doi.org/10.1121/1.1904270} {\bibfield  {journal}
  {\bibinfo  {journal} {J. Acoust. Soc. Am.}\ }\textbf {\bibinfo {volume}
  {117}},\ \bibinfo {pages} {3413} (\bibinfo {year} {2005})}\BibitemShut
  {NoStop}%
\bibitem [{\citenamefont {Martin}(2006)}]{martin2006}%
  \BibitemOpen
  \bibfield  {author} {\bibinfo {author} {\bibfnamefont {P.~A.}\ \bibnamefont
  {Martin}},\ }\href {https://doi.org/10.1017/CBO9780511735110} {\emph
  {\bibinfo {title} {Multiple Scattering: Interaction of Time-Harmonic Waves
  with {$N$}~Obstacles}}}\ (\bibinfo  {publisher} {Cambridge University
  Press},\ \bibinfo {year} {2006})\BibitemShut {NoStop}%
\bibitem [{\citenamefont {Sch{\"a}fer}\ \emph {et~al.}(2012)\citenamefont
  {Sch{\"a}fer}, \citenamefont {Lee},\ and\ \citenamefont
  {Kienle}}]{schaefer2012calculation}%
  \BibitemOpen
  \bibfield  {author} {\bibinfo {author} {\bibfnamefont {J.}~\bibnamefont
  {Sch{\"a}fer}}, \bibinfo {author} {\bibfnamefont {S.-C.}\ \bibnamefont
  {Lee}},\ and\ \bibinfo {author} {\bibfnamefont {A.}~\bibnamefont {Kienle}},\
  }\href {https://doi.org/10.1016/j.jqsrt.2012.05.019} {\bibfield  {journal}
  {\bibinfo  {journal} {J. Quant. Spectrosc. Radiat. Transfer}\ }\textbf
  {\bibinfo {volume} {113}},\ \bibinfo {pages} {2113} (\bibinfo {year}
  {2012})}\BibitemShut {NoStop}%
\bibitem [{\citenamefont {Beutel}\ \emph {et~al.}(2024)\citenamefont {Beutel},
  \citenamefont {Fernandez-Corbaton},\ and\ \citenamefont
  {Rockstuhl}}]{beutel2024treams}%
  \BibitemOpen
  \bibfield  {author} {\bibinfo {author} {\bibfnamefont {D.}~\bibnamefont
  {Beutel}}, \bibinfo {author} {\bibfnamefont {I.}~\bibnamefont
  {Fernandez-Corbaton}},\ and\ \bibinfo {author} {\bibfnamefont
  {C.}~\bibnamefont {Rockstuhl}},\ }\href
  {https://doi.org/10.1016/j.cpc.2023.109076} {\bibfield  {journal} {\bibinfo
  {journal} {Computer Physics Communications}\ }\textbf {\bibinfo {volume}
  {297}},\ \bibinfo {pages} {109076} (\bibinfo {year} {2024})}\BibitemShut
  {NoStop}%
\bibitem [{\citenamefont {Loulas}\ \emph {et~al.}(2025)\citenamefont {Loulas},
  \citenamefont {Almpanis}, \citenamefont {Kouroublakis}, \citenamefont
  {Tsakmakidis}, \citenamefont {Rockstuhl},\ and\ \citenamefont
  {Zouros}}]{2D_Loulas2025}%
  \BibitemOpen
  \bibfield  {author} {\bibinfo {author} {\bibfnamefont {I.}~\bibnamefont
  {Loulas}}, \bibinfo {author} {\bibfnamefont {E.}~\bibnamefont {Almpanis}},
  \bibinfo {author} {\bibfnamefont {M.}~\bibnamefont {Kouroublakis}}, \bibinfo
  {author} {\bibfnamefont {K.~L.}\ \bibnamefont {Tsakmakidis}}, \bibinfo
  {author} {\bibfnamefont {C.}~\bibnamefont {Rockstuhl}},\ and\ \bibinfo
  {author} {\bibfnamefont {G.~P.}\ \bibnamefont {Zouros}},\ }\href
  {https://doi.org/10.1021/acsphotonics.4c02194} {\bibfield  {journal}
  {\bibinfo  {journal} {ACS Photonics}\ }\textbf {\bibinfo {volume} {12}},\
  \bibinfo {pages} {1524} (\bibinfo {year} {2025})}\BibitemShut {NoStop}%
\bibitem [{\citenamefont {Tanaka}(2026)}]{tanaka2026calculating}%
  \BibitemOpen
  \bibfield  {author} {\bibinfo {author} {\bibfnamefont {T.}~\bibnamefont
  {Tanaka}},\ }\bibfield  {journal} {\bibinfo  {journal} {JASA Express
  Letters}\ }\textbf {\bibinfo {volume} {6}},\ \href
  {https://doi.org/10.1121/10.0043125} {10.1121/10.0043125} (\bibinfo {year}
  {2026})\BibitemShut {NoStop}%
\bibitem [{\citenamefont {Mackowski}(1994)}]{mackowski1996}%
  \BibitemOpen
  \bibfield  {author} {\bibinfo {author} {\bibfnamefont {D.~W.}\ \bibnamefont
  {Mackowski}},\ }\href {https://doi.org/10.1364/JOSAA.11.002851} {\bibfield
  {journal} {\bibinfo  {journal} {J. Opt. Soc. Am. A}\ }\textbf {\bibinfo
  {volume} {11}},\ \bibinfo {pages} {2851} (\bibinfo {year}
  {1994})}\BibitemShut {NoStop}%
\bibitem [{\citenamefont {Knight}\ \emph {et~al.}(2014)\citenamefont {Knight},
  \citenamefont {King}, \citenamefont {Liu}, \citenamefont {Everitt},
  \citenamefont {Nordlander},\ and\ \citenamefont {Halas}}]{knight2014}%
  \BibitemOpen
  \bibfield  {author} {\bibinfo {author} {\bibfnamefont {M.~W.}\ \bibnamefont
  {Knight}}, \bibinfo {author} {\bibfnamefont {N.~S.}\ \bibnamefont {King}},
  \bibinfo {author} {\bibfnamefont {L.}~\bibnamefont {Liu}}, \bibinfo {author}
  {\bibfnamefont {H.~O.}\ \bibnamefont {Everitt}}, \bibinfo {author}
  {\bibfnamefont {P.}~\bibnamefont {Nordlander}},\ and\ \bibinfo {author}
  {\bibfnamefont {N.~J.}\ \bibnamefont {Halas}},\ }\href
  {https://doi.org/10.1021/nn405495q} {\bibfield  {journal} {\bibinfo
  {journal} {ACS Nano}\ }\textbf {\bibinfo {volume} {8}},\ \bibinfo {pages}
  {834} (\bibinfo {year} {2014})}\BibitemShut {NoStop}%
\bibitem [{\citenamefont {G{\'e}rard}\ and\ \citenamefont
  {Gray}(2015)}]{gerard2015}%
  \BibitemOpen
  \bibfield  {author} {\bibinfo {author} {\bibfnamefont {D.}~\bibnamefont
  {G{\'e}rard}}\ and\ \bibinfo {author} {\bibfnamefont {S.~K.}\ \bibnamefont
  {Gray}},\ }\href {https://doi.org/10.1088/0022-3727/48/18/184001} {\bibfield
  {journal} {\bibinfo  {journal} {J. Phys. D: Appl. Phys.}\ }\textbf {\bibinfo
  {volume} {48}},\ \bibinfo {pages} {184001} (\bibinfo {year}
  {2015})}\BibitemShut {NoStop}%
\bibitem [{\citenamefont {Th{\o}gersen}\ \emph {et~al.}(2023)\citenamefont
  {Th{\o}gersen}, \citenamefont {Jensen}, \citenamefont {Belle}, \citenamefont
  {Stange}, \citenamefont {Reinertsen}, \citenamefont {Kjeldstad},
  \citenamefont {Prytz}, \citenamefont {Monakhov},\ and\ \citenamefont
  {Kepaptsoglou}}]{thogersen2023plasmonic}%
  \BibitemOpen
  \bibfield  {author} {\bibinfo {author} {\bibfnamefont {A.}~\bibnamefont
  {Th{\o}gersen}}, \bibinfo {author} {\bibfnamefont {I.~J.}\ \bibnamefont
  {Jensen}}, \bibinfo {author} {\bibfnamefont {B.~D.}\ \bibnamefont {Belle}},
  \bibinfo {author} {\bibfnamefont {M.}~\bibnamefont {Stange}}, \bibinfo
  {author} {\bibfnamefont {V.~M.}\ \bibnamefont {Reinertsen}}, \bibinfo
  {author} {\bibfnamefont {T.}~\bibnamefont {Kjeldstad}}, \bibinfo {author}
  {\bibfnamefont {{\O}.}~\bibnamefont {Prytz}}, \bibinfo {author}
  {\bibfnamefont {E.}~\bibnamefont {Monakhov}},\ and\ \bibinfo {author}
  {\bibfnamefont {D.}~\bibnamefont {Kepaptsoglou}},\ }\href
  {https://doi.org/10.1088/1361-648X/aca30e} {\bibfield  {journal} {\bibinfo
  {journal} {Journal of Physics: Condensed Matter}\ }\textbf {\bibinfo {volume}
  {35}},\ \bibinfo {pages} {065301} (\bibinfo {year} {2023})}\BibitemShut
  {NoStop}%
\bibitem [{\citenamefont {Nordlander}\ \emph {et~al.}(2004)\citenamefont
  {Nordlander}, \citenamefont {Oubre}, \citenamefont {Prodan}, \citenamefont
  {Li},\ and\ \citenamefont {Stockman}}]{nordlander2004}%
  \BibitemOpen
  \bibfield  {author} {\bibinfo {author} {\bibfnamefont {P.}~\bibnamefont
  {Nordlander}}, \bibinfo {author} {\bibfnamefont {C.}~\bibnamefont {Oubre}},
  \bibinfo {author} {\bibfnamefont {E.}~\bibnamefont {Prodan}}, \bibinfo
  {author} {\bibfnamefont {K.}~\bibnamefont {Li}},\ and\ \bibinfo {author}
  {\bibfnamefont {M.~I.}\ \bibnamefont {Stockman}},\ }\href
  {https://doi.org/10.1021/nl049681c} {\bibfield  {journal} {\bibinfo
  {journal} {Nano Lett.}\ }\textbf {\bibinfo {volume} {4}},\ \bibinfo {pages}
  {899} (\bibinfo {year} {2004})}\BibitemShut {NoStop}%
\bibitem [{\citenamefont {Prodan}\ \emph {et~al.}(2003)\citenamefont {Prodan},
  \citenamefont {Radloff}, \citenamefont {Halas},\ and\ \citenamefont
  {Nordlander}}]{prodan2003}%
  \BibitemOpen
  \bibfield  {author} {\bibinfo {author} {\bibfnamefont {E.}~\bibnamefont
  {Prodan}}, \bibinfo {author} {\bibfnamefont {C.}~\bibnamefont {Radloff}},
  \bibinfo {author} {\bibfnamefont {N.~J.}\ \bibnamefont {Halas}},\ and\
  \bibinfo {author} {\bibfnamefont {P.}~\bibnamefont {Nordlander}},\ }\href
  {https://doi.org/10.1126/science.1089171} {\bibfield  {journal} {\bibinfo
  {journal} {Science}\ }\textbf {\bibinfo {volume} {302}},\ \bibinfo {pages}
  {419} (\bibinfo {year} {2003})}\BibitemShut {NoStop}%
\bibitem [{\citenamefont {Brandl}\ \emph {et~al.}(2006)\citenamefont {Brandl},
  \citenamefont {Mirin},\ and\ \citenamefont {Nordlander}}]{brandl2006}%
  \BibitemOpen
  \bibfield  {author} {\bibinfo {author} {\bibfnamefont {D.~W.}\ \bibnamefont
  {Brandl}}, \bibinfo {author} {\bibfnamefont {N.~A.}\ \bibnamefont {Mirin}},\
  and\ \bibinfo {author} {\bibfnamefont {P.}~\bibnamefont {Nordlander}},\
  }\href {https://doi.org/10.1021/jp0613485} {\bibfield  {journal} {\bibinfo
  {journal} {J. Phys. Chem. B}\ }\textbf {\bibinfo {volume} {110}},\ \bibinfo
  {pages} {12302} (\bibinfo {year} {2006})}\BibitemShut {NoStop}%
\bibitem [{\citenamefont {Alegret}\ \emph {et~al.}(2008)\citenamefont
  {Alegret}, \citenamefont {Rindzevicius}, \citenamefont {Pakizeh},
  \citenamefont {Alaverdyan}, \citenamefont {Gunnarsson},\ and\ \citenamefont
  {K{\"a}ll}}]{alegret2008}%
  \BibitemOpen
  \bibfield  {author} {\bibinfo {author} {\bibfnamefont {J.}~\bibnamefont
  {Alegret}}, \bibinfo {author} {\bibfnamefont {T.}~\bibnamefont
  {Rindzevicius}}, \bibinfo {author} {\bibfnamefont {T.}~\bibnamefont
  {Pakizeh}}, \bibinfo {author} {\bibfnamefont {Y.}~\bibnamefont {Alaverdyan}},
  \bibinfo {author} {\bibfnamefont {L.}~\bibnamefont {Gunnarsson}},\ and\
  \bibinfo {author} {\bibfnamefont {M.}~\bibnamefont {K{\"a}ll}},\ }\href
  {https://doi.org/10.1021/jp804505k} {\bibfield  {journal} {\bibinfo
  {journal} {J. Phys. Chem. C}\ }\textbf {\bibinfo {volume} {112}},\ \bibinfo
  {pages} {14313} (\bibinfo {year} {2008})}\BibitemShut {NoStop}%
\bibitem [{\citenamefont {Zhao}\ \emph {et~al.}(2024)\citenamefont {Zhao},
  \citenamefont {Liu}, \citenamefont {Chen}, \citenamefont {Shi}, \citenamefont
  {Li}, \citenamefont {Tang}, \citenamefont {Zhu}, \citenamefont {Li},
  \citenamefont {Yao}, \citenamefont {Wei}, \citenamefont {Song}, \citenamefont
  {Sun}, \citenamefont {Fan}, \citenamefont {Zhou}, \citenamefont {Qiu},\ and\
  \citenamefont {Hao}}]{Zhao2024}%
  \BibitemOpen
  \bibfield  {author} {\bibinfo {author} {\bibfnamefont {X.}~\bibnamefont
  {Zhao}}, \bibinfo {author} {\bibfnamefont {X.}~\bibnamefont {Liu}}, \bibinfo
  {author} {\bibfnamefont {D.}~\bibnamefont {Chen}}, \bibinfo {author}
  {\bibfnamefont {G.}~\bibnamefont {Shi}}, \bibinfo {author} {\bibfnamefont
  {G.}~\bibnamefont {Li}}, \bibinfo {author} {\bibfnamefont {X.}~\bibnamefont
  {Tang}}, \bibinfo {author} {\bibfnamefont {X.}~\bibnamefont {Zhu}}, \bibinfo
  {author} {\bibfnamefont {M.}~\bibnamefont {Li}}, \bibinfo {author}
  {\bibfnamefont {L.}~\bibnamefont {Yao}}, \bibinfo {author} {\bibfnamefont
  {Y.}~\bibnamefont {Wei}}, \bibinfo {author} {\bibfnamefont {W.}~\bibnamefont
  {Song}}, \bibinfo {author} {\bibfnamefont {Z.}~\bibnamefont {Sun}}, \bibinfo
  {author} {\bibfnamefont {X.}~\bibnamefont {Fan}}, \bibinfo {author}
  {\bibfnamefont {Z.}~\bibnamefont {Zhou}}, \bibinfo {author} {\bibfnamefont
  {T.}~\bibnamefont {Qiu}},\ and\ \bibinfo {author} {\bibfnamefont
  {Q.}~\bibnamefont {Hao}},\ }\bibfield  {journal} {\bibinfo  {journal} {Nature
  Communications}\ }\textbf {\bibinfo {volume} {15}},\ \href
  {https://doi.org/10.1038/s41467-024-50321-0} {10.1038/s41467-024-50321-0}
  (\bibinfo {year} {2024})\BibitemShut {NoStop}%
\bibitem [{\citenamefont {Landau}\ \emph {et~al.}(2013)\citenamefont {Landau},
  \citenamefont {Bell}, \citenamefont {Kearsley}, \citenamefont {Pitaevskii},
  \citenamefont {Lifshitz},\ and\ \citenamefont {Sykes}}]{LL_Electrodyn}%
  \BibitemOpen
  \bibfield  {author} {\bibinfo {author} {\bibfnamefont {L.~D.}\ \bibnamefont
  {Landau}}, \bibinfo {author} {\bibfnamefont {J.~S.}\ \bibnamefont {Bell}},
  \bibinfo {author} {\bibfnamefont {M.}~\bibnamefont {Kearsley}}, \bibinfo
  {author} {\bibfnamefont {L.}~\bibnamefont {Pitaevskii}}, \bibinfo {author}
  {\bibfnamefont {E.}~\bibnamefont {Lifshitz}},\ and\ \bibinfo {author}
  {\bibfnamefont {J.}~\bibnamefont {Sykes}},\ }\href@noop {} {\emph {\bibinfo
  {title} {Electrodynamics of continuous media}}},\ Vol.~\bibinfo {volume} {8}\
  (\bibinfo  {publisher} {elsevier},\ \bibinfo {year} {2013})\BibitemShut
  {NoStop}%
\bibitem [{\citenamefont {Lei}\ \emph {et~al.}(2010)\citenamefont {Lei},
  \citenamefont {Aubry}, \citenamefont {Maier},\ and\ \citenamefont
  {Pendry}}]{lei2010broadband}%
  \BibitemOpen
  \bibfield  {author} {\bibinfo {author} {\bibfnamefont {D.~Y.}\ \bibnamefont
  {Lei}}, \bibinfo {author} {\bibfnamefont {A.}~\bibnamefont {Aubry}}, \bibinfo
  {author} {\bibfnamefont {S.~A.}\ \bibnamefont {Maier}},\ and\ \bibinfo
  {author} {\bibfnamefont {J.~B.}\ \bibnamefont {Pendry}},\ }\href
  {https://doi.org/10.1088/1367-2630/12/9/093030} {\bibfield  {journal}
  {\bibinfo  {journal} {New Journal of Physics}\ }\textbf {\bibinfo {volume}
  {12}},\ \bibinfo {pages} {093030} (\bibinfo {year} {2010})}\BibitemShut
  {NoStop}%
\bibitem [{\citenamefont {Tribelsky}(2023)}]{Tribelsky2023}%
  \BibitemOpen
  \bibfield  {author} {\bibinfo {author} {\bibfnamefont {M.~I.}\ \bibnamefont
  {Tribelsky}},\ }\bibfield  {journal} {\bibinfo  {journal} {Laser \& Photonics
  Reviews}\ }\textbf {\bibinfo {volume} {18}},\ \href
  {https://doi.org/10.1002/lpor.202300512} {10.1002/lpor.202300512} (\bibinfo
  {year} {2023})\BibitemShut {NoStop}%
\bibitem [{\citenamefont {Abramowitz}\ and\ \citenamefont
  {Stegun}(1964)}]{abramowitz1964}%
  \BibitemOpen
  \bibfield  {author} {\bibinfo {author} {\bibfnamefont {M.}~\bibnamefont
  {Abramowitz}}\ and\ \bibinfo {author} {\bibfnamefont {I.~A.}\ \bibnamefont
  {Stegun}},\ }\href@noop {} {\emph {\bibinfo {title} {Handbook of Mathematical
  Functions}}}\ (\bibinfo  {publisher} {Dover},\ \bibinfo {address} {New
  York},\ \bibinfo {year} {1964})\BibitemShut {NoStop}%
\bibitem [{\citenamefont {Watson}(1944)}]{watson1944}%
  \BibitemOpen
  \bibfield  {author} {\bibinfo {author} {\bibfnamefont {G.~N.}\ \bibnamefont
  {Watson}},\ }\href@noop {} {\emph {\bibinfo {title} {A Treatise on the Theory
  of Bessel Functions}}},\ \bibinfo {edition} {2nd}\ ed.\ (\bibinfo
  {publisher} {Cambridge University Press},\ \bibinfo {year}
  {1944})\BibitemShut {NoStop}%
\bibitem [{\citenamefont {Hulst}(2018)}]{Hulst2018}%
  \BibitemOpen
  \bibfield  {author} {\bibinfo {author} {\bibfnamefont {H.~C.}\ \bibnamefont
  {Hulst}},\ }\href@noop {} {\emph {\bibinfo {title} {Light scattering by small
  particles}}},\ \bibinfo {edition} {unabridged and corrected republication of
  the work originally published in 1957 by \protect{John Wiley and Sons, Inc.,
  N.Y.}}\ ed.,\ Dover books on physics\ (\bibinfo  {publisher} {Dover
  Publications, Inc.},\ \bibinfo {address} {New York},\ \bibinfo {year}
  {2018})\BibitemShut {NoStop}%
\bibitem [{\citenamefont {Higham}(2002)}]{higham2002accuracy}%
  \BibitemOpen
  \bibfield  {author} {\bibinfo {author} {\bibfnamefont {N.~J.}\ \bibnamefont
  {Higham}},\ }\href@noop {} {\emph {\bibinfo {title} {Accuracy and Stability
  of Numerical Algorithms}}},\ \bibinfo {edition} {2nd}\ ed.\ (\bibinfo
  {publisher} {SIAM},\ \bibinfo {address} {Philadelphia},\ \bibinfo {year}
  {2002})\BibitemShut {NoStop}%
\bibitem [{\citenamefont {Harris}\ \emph {et~al.}(2020)\citenamefont {Harris},
  \citenamefont {Millman}, \citenamefont {van~der Walt} \emph
  {et~al.}}]{harris2020numpy}%
  \BibitemOpen
  \bibfield  {author} {\bibinfo {author} {\bibfnamefont {C.~R.}\ \bibnamefont
  {Harris}}, \bibinfo {author} {\bibfnamefont {K.~J.}\ \bibnamefont {Millman}},
  \bibinfo {author} {\bibfnamefont {S.~J.}\ \bibnamefont {van~der Walt}}, \emph
  {et~al.},\ }\href {https://doi.org/10.1038/s41586-020-2649-2} {\bibfield
  {journal} {\bibinfo  {journal} {Nature}\ }\textbf {\bibinfo {volume} {585}},\
  \bibinfo {pages} {357} (\bibinfo {year} {2020})}\BibitemShut {NoStop}%
\bibitem [{\citenamefont {Johansson}(2013)}]{johansson2013mpmath}%
  \BibitemOpen
  \bibfield  {author} {\bibinfo {author} {\bibfnamefont {F.}~\bibnamefont
  {Johansson}},\ }\href@noop {} {\  (\bibinfo {year} {2013})},\ \bibinfo {note}
  {\url{http://mpmath.org/}}\BibitemShut {NoStop}%
\bibitem [{\citenamefont {Anderson}\ \emph {et~al.}(1999)\citenamefont
  {Anderson}, \citenamefont {Bai}, \citenamefont {Bischof}, \citenamefont
  {Blackford}, \citenamefont {Demmel}, \citenamefont {Dongarra}, \citenamefont
  {Du~Croz}, \citenamefont {Greenbaum}, \citenamefont {Hammarling},
  \citenamefont {McKenney},\ and\ \citenamefont
  {Sorensen}}]{anderson1999lapack}%
  \BibitemOpen
  \bibfield  {author} {\bibinfo {author} {\bibfnamefont {E.}~\bibnamefont
  {Anderson}}, \bibinfo {author} {\bibfnamefont {Z.}~\bibnamefont {Bai}},
  \bibinfo {author} {\bibfnamefont {C.}~\bibnamefont {Bischof}}, \bibinfo
  {author} {\bibfnamefont {S.}~\bibnamefont {Blackford}}, \bibinfo {author}
  {\bibfnamefont {J.}~\bibnamefont {Demmel}}, \bibinfo {author} {\bibfnamefont
  {J.}~\bibnamefont {Dongarra}}, \bibinfo {author} {\bibfnamefont
  {J.}~\bibnamefont {Du~Croz}}, \bibinfo {author} {\bibfnamefont
  {A.}~\bibnamefont {Greenbaum}}, \bibinfo {author} {\bibfnamefont
  {S.}~\bibnamefont {Hammarling}}, \bibinfo {author} {\bibfnamefont
  {A.}~\bibnamefont {McKenney}},\ and\ \bibinfo {author} {\bibfnamefont
  {D.}~\bibnamefont {Sorensen}},\ }\href@noop {} {\emph {\bibinfo {title}
  {{LAPACK} Users' Guide}}},\ \bibinfo {edition} {3rd}\ ed.\ (\bibinfo
  {publisher} {SIAM},\ \bibinfo {address} {Philadelphia},\ \bibinfo {year}
  {1999})\BibitemShut {NoStop}%
\bibitem [{\citenamefont {Meurer}\ \emph {et~al.}(2017)\citenamefont {Meurer},
  \citenamefont {Smith}, \citenamefont {Paprocki} \emph
  {et~al.}}]{meurer2017sympy}%
  \BibitemOpen
  \bibfield  {author} {\bibinfo {author} {\bibfnamefont {A.}~\bibnamefont
  {Meurer}}, \bibinfo {author} {\bibfnamefont {C.~P.}\ \bibnamefont {Smith}},
  \bibinfo {author} {\bibfnamefont {M.}~\bibnamefont {Paprocki}}, \emph
  {et~al.},\ }\href {https://doi.org/10.7717/peerj-cs.103} {\bibfield
  {journal} {\bibinfo  {journal} {PeerJ Comput. Sci.}\ }\textbf {\bibinfo
  {volume} {3}},\ \bibinfo {pages} {e103} (\bibinfo {year} {2017})}\BibitemShut
  {NoStop}%
\bibitem [{\citenamefont {Bogacki}\ and\ \citenamefont
  {Shampine}(1989)}]{bogacki1989pair}%
  \BibitemOpen
  \bibfield  {author} {\bibinfo {author} {\bibfnamefont {P.}~\bibnamefont
  {Bogacki}}\ and\ \bibinfo {author} {\bibfnamefont {L.~F.}\ \bibnamefont
  {Shampine}},\ }\href {https://doi.org/10.1016/0893-9659(89)90079-7}
  {\bibfield  {journal} {\bibinfo  {journal} {Appl. Math. Lett.}\ }\textbf
  {\bibinfo {volume} {2}},\ \bibinfo {pages} {321} (\bibinfo {year}
  {1989})}\BibitemShut {NoStop}%
\bibitem [{\citenamefont {Hairer}\ \emph {et~al.}(1993)\citenamefont {Hairer},
  \citenamefont {N{\o}rsett},\ and\ \citenamefont
  {Wanner}}]{hairer1993solving}%
  \BibitemOpen
  \bibfield  {author} {\bibinfo {author} {\bibfnamefont {E.}~\bibnamefont
  {Hairer}}, \bibinfo {author} {\bibfnamefont {S.~P.}\ \bibnamefont
  {N{\o}rsett}},\ and\ \bibinfo {author} {\bibfnamefont {G.}~\bibnamefont
  {Wanner}},\ }\href@noop {} {\emph {\bibinfo {title} {Solving Ordinary
  Differential Equations {I}: Nonstiff Problems}}},\ \bibinfo {edition} {2nd}\
  ed.\ (\bibinfo  {publisher} {Springer},\ \bibinfo {address} {Berlin},\
  \bibinfo {year} {1993})\BibitemShut {NoStop}%
\bibitem [{\citenamefont {Palik}(1998)}]{palik1998}%
  \BibitemOpen
  \bibfield  {author} {\bibinfo {author} {\bibfnamefont {E.~D.}\ \bibnamefont
  {Palik}},\ }\href@noop {} {\emph {\bibinfo {title} {Handbook of Optical
  Constants of Solids}}}\ (\bibinfo  {publisher} {Academic Press},\ \bibinfo
  {address} {San Diego},\ \bibinfo {year} {1998})\BibitemShut {NoStop}%
\bibitem [{\citenamefont {Molesky}\ \emph {et~al.}(2018)\citenamefont
  {Molesky}, \citenamefont {Lin}, \citenamefont {Piggott}, \citenamefont {Jin},
  \citenamefont {Vuckovi{\'c}},\ and\ \citenamefont {Rodriguez}}]{molesky2018}%
  \BibitemOpen
  \bibfield  {author} {\bibinfo {author} {\bibfnamefont {S.}~\bibnamefont
  {Molesky}}, \bibinfo {author} {\bibfnamefont {Z.}~\bibnamefont {Lin}},
  \bibinfo {author} {\bibfnamefont {A.~Y.}\ \bibnamefont {Piggott}}, \bibinfo
  {author} {\bibfnamefont {W.}~\bibnamefont {Jin}}, \bibinfo {author}
  {\bibfnamefont {J.}~\bibnamefont {Vuckovi{\'c}}},\ and\ \bibinfo {author}
  {\bibfnamefont {A.~W.}\ \bibnamefont {Rodriguez}},\ }\href
  {https://doi.org/10.1038/s41566-018-0246-9} {\bibfield  {journal} {\bibinfo
  {journal} {Nature Photon.}\ }\textbf {\bibinfo {volume} {12}},\ \bibinfo
  {pages} {659} (\bibinfo {year} {2018})}\BibitemShut {NoStop}%
\bibitem [{\citenamefont {Hao}\ and\ \citenamefont {Schatz}(2004)}]{hao2004}%
  \BibitemOpen
  \bibfield  {author} {\bibinfo {author} {\bibfnamefont {E.}~\bibnamefont
  {Hao}}\ and\ \bibinfo {author} {\bibfnamefont {G.~C.}\ \bibnamefont
  {Schatz}},\ }\href {https://doi.org/10.1063/1.1629280} {\bibfield  {journal}
  {\bibinfo  {journal} {J. Chem. Phys.}\ }\textbf {\bibinfo {volume} {120}},\
  \bibinfo {pages} {357} (\bibinfo {year} {2004})}\BibitemShut {NoStop}%
\bibitem [{\citenamefont {Camden}\ \emph {et~al.}(2008)\citenamefont {Camden},
  \citenamefont {Dieringer}, \citenamefont {Wang}, \citenamefont {Masiello},
  \citenamefont {Marks}, \citenamefont {Schatz},\ and\ \citenamefont
  {Van~Duyne}}]{camden2008}%
  \BibitemOpen
  \bibfield  {author} {\bibinfo {author} {\bibfnamefont {J.~P.}\ \bibnamefont
  {Camden}}, \bibinfo {author} {\bibfnamefont {J.~A.}\ \bibnamefont
  {Dieringer}}, \bibinfo {author} {\bibfnamefont {Y.}~\bibnamefont {Wang}},
  \bibinfo {author} {\bibfnamefont {D.~J.}\ \bibnamefont {Masiello}}, \bibinfo
  {author} {\bibfnamefont {L.~D.}\ \bibnamefont {Marks}}, \bibinfo {author}
  {\bibfnamefont {G.~C.}\ \bibnamefont {Schatz}},\ and\ \bibinfo {author}
  {\bibfnamefont {R.~P.}\ \bibnamefont {Van~Duyne}},\ }\href
  {https://doi.org/10.1021/ja8051427} {\bibfield  {journal} {\bibinfo
  {journal} {J. Am. Chem. Soc.}\ }\textbf {\bibinfo {volume} {130}},\ \bibinfo
  {pages} {12616} (\bibinfo {year} {2008})}\BibitemShut {NoStop}%
\bibitem [{\citenamefont {Jin}(2015)}]{jin2015}%
  \BibitemOpen
  \bibfield  {author} {\bibinfo {author} {\bibfnamefont {J.-M.}\ \bibnamefont
  {Jin}},\ }\href@noop {} {\emph {\bibinfo {title} {The Finite Element Method
  in Electromagnetics}}},\ \bibinfo {edition} {3rd}\ ed.\ (\bibinfo
  {publisher} {Wiley-IEEE Press},\ \bibinfo {address} {Hoboken, NJ},\ \bibinfo
  {year} {2015})\BibitemShut {NoStop}%
\bibitem [{\citenamefont {Monk}(2003)}]{monk2003}%
  \BibitemOpen
  \bibfield  {author} {\bibinfo {author} {\bibfnamefont {P.}~\bibnamefont
  {Monk}},\ }\href {https://doi.org/10.1093/acprof:oso/9780198508885.001.0001}
  {\emph {\bibinfo {title} {Finite Element Methods for {Maxwell's}
  Equations}}}\ (\bibinfo  {publisher} {Oxford University Press},\ \bibinfo
  {address} {Oxford},\ \bibinfo {year} {2003})\BibitemShut {NoStop}%
\bibitem [{\citenamefont {Taflove}\ and\ \citenamefont
  {Hagness}(2005)}]{taflove2005}%
  \BibitemOpen
  \bibfield  {author} {\bibinfo {author} {\bibfnamefont {A.}~\bibnamefont
  {Taflove}}\ and\ \bibinfo {author} {\bibfnamefont {S.~C.}\ \bibnamefont
  {Hagness}},\ }\href@noop {} {\emph {\bibinfo {title} {Computational
  Electrodynamics: The Finite-Difference Time-Domain Method}}},\ \bibinfo
  {edition} {3rd}\ ed.\ (\bibinfo  {publisher} {Artech House},\ \bibinfo
  {address} {Boston},\ \bibinfo {year} {2005})\BibitemShut {NoStop}%
\bibitem [{\citenamefont {Yee}(1966)}]{yee1966}%
  \BibitemOpen
  \bibfield  {author} {\bibinfo {author} {\bibfnamefont {K.~S.}\ \bibnamefont
  {Yee}},\ }\href {https://doi.org/10.1109/TAP.1966.1138693} {\bibfield
  {journal} {\bibinfo  {journal} {IEEE Trans. Antennas Propag.}\ }\textbf
  {\bibinfo {volume} {14}},\ \bibinfo {pages} {302} (\bibinfo {year}
  {1966})}\BibitemShut {NoStop}%
\end{thebibliography}%

\end{document}